\UseRawInputEncoding

\documentclass[journal]{IEEEtran}
\usepackage[justification=centering]{caption}
\usepackage{lineno,hyperref}
\usepackage{amsmath,graphicx}
\usepackage{epsfig}
\usepackage{amsfonts,amssymb}
\usepackage{bm}
\usepackage{hyperref}
\usepackage{array}
\usepackage{enumerate,multirow}
\usepackage{subfigure}
\usepackage{algorithm}
\usepackage{algorithmic}
\usepackage{xcolor}
\usepackage{float}
\usepackage{makecell}
\usepackage{booktabs}
\usepackage{bbding}
\usepackage{textcomp}
\usepackage{cite}
\usepackage{booktabs}
\usepackage{tabularx}
\usepackage{multirow}
\usepackage{graphicx}
\usepackage{bbm}

\begin{document}

\title{Information Fusion in Attention Networks Using Adaptive and Multi-level Factorized Bilinear Pooling for Audio-visual Emotion Recognition}

\author{Hengshun Zhou, Jun Du, Yuanyuan Zhang, Qing Wang, Qing-Feng Liu, and Chin-Hui Lee,~\IEEEmembership{Fellow,~IEEE}
\thanks{H. Zhou, J. Du, Y. Zhang, Q. Wang and Q.-F. Liu are with the National Engineering Laboratory for Speech and Language Information Processing, University of Science and Technology of China, Hefei 230027, China (e-mail: zhhs@mail.ustc.edu.cn; jundu@ustc.edu.cn; zyuan@mail.ustc.edu.cn; xiaosong@mail.ustc.edu.cn; qfliu@iflytek.com).}\thanks{C.-H. Lee is with the School of Electrical and Computer Engineering, Georgia Institute of Technology, Atlanta, GA. 30332-0250, USA (e-mail: chl@ece.gatech.edu).}}

\markboth{IEEE/ACM TRANSACTIONS ON AUDIO, SPEECH, AND LANGUAGE PROCESSING}%
{Shell \MakeLowercase{\textit{et al.}}: Bare Demo of IEEEtran.cls for IEEE Journals}

\maketitle
\begin{abstract}
Multimodal emotion recognition is a challenging task in emotion computing as it is quite difficult to extract discriminative features to identify the subtle differences in human emotions with abstract concept and multiple expressions. Moreover, how to fully utilize both audio and visual information is still an open problem. In this paper, we propose a novel multimodal fusion attention network for audio-visual emotion recognition based on adaptive and multi-level factorized bilinear pooling (FBP). First, for the audio stream, a fully convolutional network (FCN) equipped with 1-D attention mechanism and local response normalization is designed for speech emotion recognition. Next, a global FBP (G-FBP) approach is presented to perform audio-visual information fusion by integrating self-attention based video stream with the proposed audio stream. To improve G-FBP, an adaptive strategy (AG-FBP) to dynamically calculate the fusion weight of two modalities is devised based on the emotion-related representation vectors from the attention mechanism of respective modalities. Finally, to fully utilize the local emotion information, adaptive and multi-level FBP (AM-FBP) is introduced by combining both global-trunk and intra-trunk data in one recording on top of AG-FBP. Tested on the IEMOCAP corpus for speech emotion recognition with only audio stream, the new FCN method outperforms the state-of-the-art results with an accuracy of 71.40\%. Moreover, validated on the AFEW database of EmotiW2019 sub-challenge and the IEMOCAP corpus for audio-visual emotion recognition, the proposed AM-FBP approach achieves the best accuracy of 63.09\% and 75.49\% respectively on the test set .
\end{abstract}


\IEEEpeerreviewmaketitle

\section{Introduction}
\IEEEPARstart{D}{ue} to the rapid development of intelligent technology and the wide applications in human\verb|-|computer interaction, it is of great significance to realize scientific emotion recognition~\cite{ren2019multi-modal}. For example, doctors utilize emotion recognition technology to do research on Parkinson's disease~\cite{bowers2006faces,yuvaraj2014detection}. In the field of service robots, effective emotion recognition can bring more comfortable interaction experience to users. As for the areas of computing advertising and entertainment, detecting consumers' emotions helps enterprises provide better services~\cite{vinola2015a,satt2017efficient}. Accordingly, automatic multimodal emotion recognition, has attracted more and more attention in real-life cases~\cite{dhall2018emotiw,dhall2019emotiw}, which is also the focus of this study.

Some progress of multimodal emotion recognition has been made by combining different modalities such as speech, face, body gesture, and brain signals, etc~\cite{gunes2007bi,yuvaraj2014detection,bong2017implementation}. Audio and video, more specifically, the speech and facial expressions are two kinds of most powerful, natural and universal signals for human beings to convey their emotional states and intentions~\cite{tian2001recognizing}. For speech emotion recognition (SER) using only audio stream, distinguishing acoustic features~\cite{chen20183,zhang2018attention} are often extracted from original raw speech signals, followed by different classifiers~\cite{8470342,li2018attention}. For video-based emotion recognition, video frame or image preprocessing~\cite{meng2019frame} is necessary correspondingly, referring to face detection, alignment and face key point detection and so on. Then image feature vectors are also fed into a classifier for prediction~\cite{fan2018video-based}. As for integrating audio and video modalities, it usually involves at the feature, model and decision levels. In feature-level fusion, features extracted from each of the two modalities are concatenated as one vector for emotion classification~\cite{55503fc645ce0a409eb301e1}, which does not take into account the differences in modal-specific emotional characteristics. Moreover, this strategy is difficult to model the time synchronization between audio and visual modalities. For decision-level fusion, the posterior probabilities of the two individual classifiers are combined, e.g., using linear weighted combination, support vector machine (SVM), etc.~\cite{dobrivsek2013towards} to obtain the final recognition results. This technique fully considers the differences of audio and visual features, but it is weak in modeling the interactions between the two modalities. As a compromise between feature-level and decision-level fusions, model-level fusion has also been used for audio-visual emotion recognition (AVER). A tripled hidden Markov model (THMM) was introduced to perform the recognition which allowed the state asynchrony of the audio and visual observation sequences while preserving their natural correlation over time~\cite{song2004audio}. In~\cite{zeng2006training}, multi-stream hidden Markov model (MFHMM) was proposed which adopted a variety of learning methods to achieve a robust multi-stream fusion result according to the maximum entropy principle.

\begin{figure*}
  \centering
   \includegraphics[width=180mm]{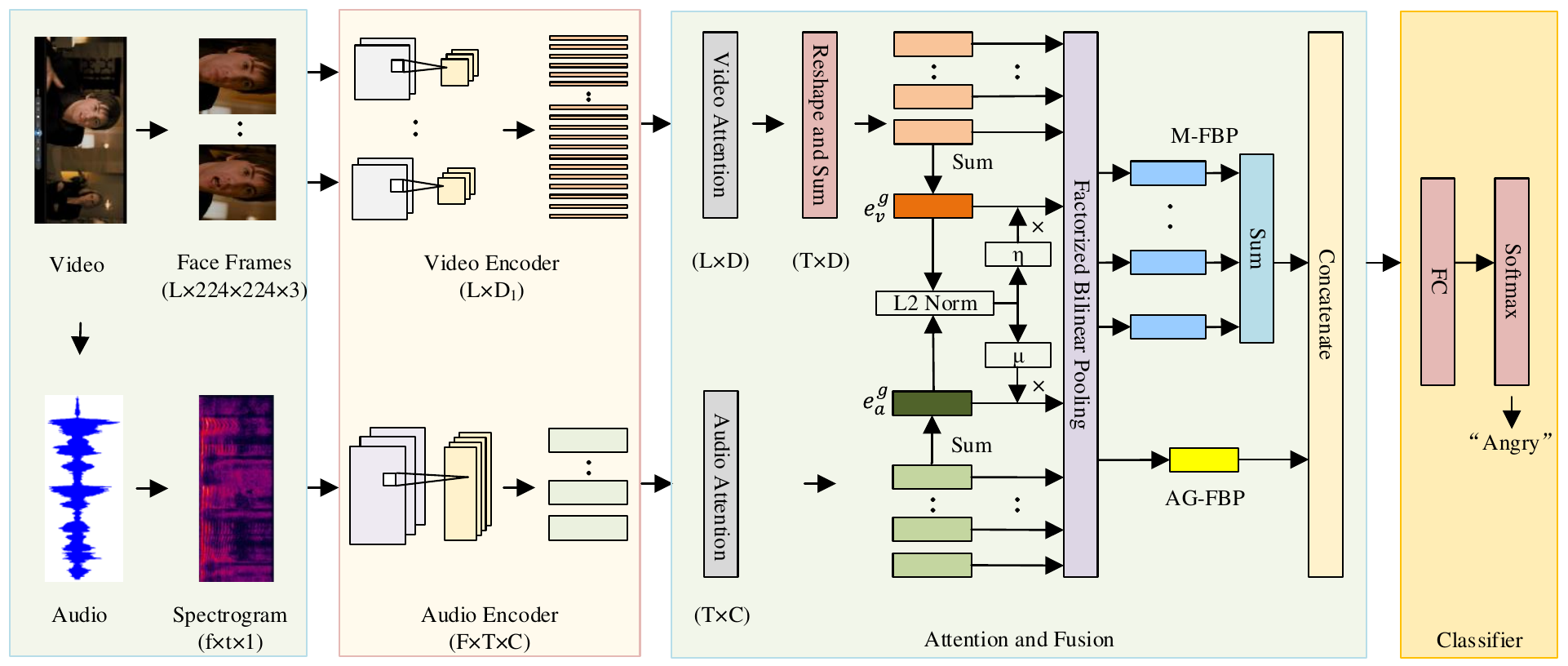}
     \caption{An overall architecture of the proposed multimodal attention and fusion network based on adaptive and multi-level factorized bilinear pooling for audio-visual emotion recognition. The detected face and spectrogram are initially encoded by video encoder and audio encoder respectively. The outputs are subsequently weighted by 1-D attention. Then AG-FBP and M-FBP are employed to fuse audio-visual information based on whole sentence and segmentation respectively. FBP module \leftline{ is shown in Fig.~\ref{fbpmodule}. The outputs are finally concatenated and fed into classifier to determine the emotion class.}}
     \label{pipeline}
\end{figure*}

With recent developments of deep learning in the multimodal field, recurrent neural network (RNN) and 3D convolutional networks (C3D) were used to solve the problem of video classification in~\cite{fan2016video}. It encodes appearance and motion information in different ways and combines them into a late-fusion manner. Researchers have also investigated how to extract more representative features~\cite{li2019bi,zhou2019exploring} as expression forms of audio and video are often quite different. Deep learning shows better performances than other traditional machine learning algorithms in this kind of fusion.~\cite{zhang2017learning} proposed to bridge the emotional gap by using a hybrid deep model, which first produces audio\verb|-|visual segment features with convolutional neural networks (CNNs) and 3D-CNN, then fuses them in deep belief networks (DBNs). In~\cite{li2019attentive}, a concatenation of different modalities was performed after an encoder which yielded significant improvements. In our recent work~\cite{zhang2019deep}, we introduced global-trunk based factorized bilinear pooling (G-FBP) to integrate the audio and visual features, achieving a state-of-the-art performance.

Audio-visual emotion recognition has been investigated for quite a few years and considered as a comparatively hot topic in the field of affective computing. Nonetheless, it remains a challenging problem in which there are still many uncontrolled factors for data acquisition. The varying conditions for audio-visual emotion data include indoor and outdoor scenarios, environmental noises, lighting situations, motion blurs, occlusions and pose changes, etc~\cite{2018Audiovisual}. To address these issues, the EmotiW~\cite{dhall2019emotiw} challenges have been held successfully since 2013. The winning teams~\cite{hu2017learning,liu2018multi,wang2019multi,zhou2019exploring} have proposed several advanced techniques for AVER and achieved better results every year, which further investigated on how to effectively model different modalities for information fusion. In this study, we comprehensively extend our previous G-FBP approach~\cite{zhang2019deep} and propose an attention network for multimodal fusion for AVER based on adaptive and multi-level FBP as shown in Fig.~\ref{pipeline}. The new contributions can be summarized below:

\begin{itemize}
 \item A fully convolutional network (FCN) based 1-D attention network is designed for speech emotion recognition by utilizing local response normalization (LRN).
 \item An adaptive G-FBP (AG-FBP) approach is presented to automatically calculate the importance weights of audio and video modalities when using G-FBP fusion.
 \item On top of AG-FBP, adaptive and multi-level FBP (AM-FBP) is introduced to fully utilize the local emotion information by additionally using intra-trunk data.
 \item We achieve the best accuracy of 63.09\% on the test set of EmotiW2019 sub-challenge \cite{dhall2019emotiw} and 75.49\% on the test set of IEMOCAP corpus, and demonstrate the effectiveness of the proposed approach by visualizing the changes of attention weights and network embedding.
\end{itemize}

The rest of this paper is organized as follows. In Section~\ref{realted_work}, we introduce the related work. In Section~\ref{proposed_method}, we elaborate on the proposed fusion strategy. In Section~\ref{exps_analysis}, we present our experimental results and analyses. Finally, we draw our conclusions in Section~\ref{conclusion}.

\section{Related Work}
\label{realted_work}

\subsection{Audio-based emotion recognition}
In the process of human interaction, speech is the most direct communication channel. People can often clearly feel the changes in emotion through speech, such as human voice quality, rhythm, as well as prosodic expressions in pitch and energy contours. In order to recognize speakers' emotional states, distinguishing paralinguistic features, which do not depend on the the lexical content, need to be extracted from speech~\cite{guidi2015automatic}. Many types of acoustic features have been used for speech emotion recognition, including continuous, qualitative, and spectral features~\cite{huang2015extraction,zhao2019speech}.

\begin{figure}
  \setlength{\belowcaptionskip}{-0.3cm}
  \centering
   \includegraphics[width=65mm]{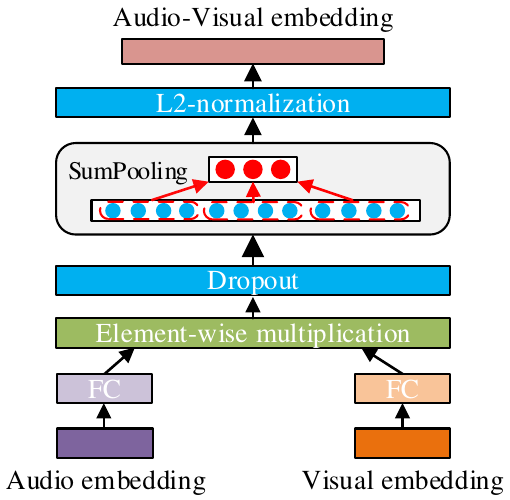}
     \caption{Factorized bilinear pooling (FBP) module.}
     \label{fbpmodule}
\end{figure}

In general, the original raw speech signals are first segmented into overlapped frames. Then various statistical functions (e.g. mean, max, linear regression coefficients, etc.) are utilized to obtain frame-level features. The outputs are then concatenated as a feature vector to represent the whole audio recording, followed with the classifiers~\cite{zhang2018attention}. As deep learning came to prominence, deep neural networks (DNNs) were utilized on top of the traditional utterance-level features and achieved a significant accuracy improvement comparing with conventional classifiers.~\cite{sarma2018emotion} investigated a number of DNN architectures and interleaved time-delay neural network and long short term memory (TDNN-LSTM) with time-restricted self-attention, achieving a good performance gain.~\cite{satt2017efficient} presented a new implementation of emotion recognition using spectrogram features and classifiers based on CNNs and RNNs. In~\cite{chen20183}, a 3-D attention-based convolution RNN (ACRNN) was proposed. An attention pooling based representation learning method was introduced in~\cite{li2018attention}. They all processed the whole speech utterance into small segments and used attention mechanisms for speech emotion recognition. In~\cite{huang2018speech}, a triplet loss was used to reinforce emotional clustering based on LSTM and explore three different strategies to handle variable-length inputs for SER. Recently, more research efforts focused on auxiliary information and innovative ways to assist emotion recognition. For example, transcripts, language cues and cross-culture information were adopted in emotion recognition~\cite{li2019attentive,liang2019cross,xu2019learning}. In~\cite{chatziagapi2019data}, conditioned data augmentation using generative adversarial networks (GANs) was explored to address the problem of data imbalance in SER tasks. Furthermore,~\cite{zhang2019attention} used multi-task learning with attention mechanism to share useful information in SER scenarios. Also, there were no agreements on appropriate features for SER. In~\cite{latif2019direct} representations were learnt from raw speech and in~\cite{sarma2019improving} phone posteriors in raw speech waveform were employed to improve emotion identification.

\subsection{Video-based emotion recognition}

In interpersonal interactions, people can enhance communication effectiveness by controlling their facial expressions, an important way to spread human emotional information. It refers to all kinds of emotions expressed through the changes of muscle movements in face, eye and mouth. Among them, the muscle groups near the eyes and mouth are shown to be the most prominent~\cite{articlefacs}. In recent years, research on visual recognition paid more attentions to feature learning via neural networks.~\cite{knyazev2018leveraging} utilized CNNs for feature extraction of facial expression recognition (FER). The winners in the AVER task of EmotiW Challenge used facial features extracted from deep CNNs trained on large face datasets~\cite{knyazev2018leveraging,liu2018multi}. In~\cite{meng2019frame}, spatial-temporal techniques aimed to model the temporal or motion information in videos. Deep C3D was a widely-used spatial-temporal approach to video-based FER~\cite{tran2015learning}. In~\cite{yang2018geometry}, geometry-based FER was proposed to boost the accuracy by using a multi-kernel framework to combine features. Effective emotion recognition was implemented in~\cite{choi2018recognizing} by learning the proposed 2D landmark information on a CNN and a LSTM-based network. In~\cite{song2019facial} and~\cite{bai2019disentangled} emotion expressions were recognized by using the differences (or relations) between neutral and expressive faces. Finally, continuous emotion recognition in videos was implemented in~\cite{wu2019continuous} by fusing facial expression, head pose and eye gaze.

\subsection{Audio-visual based emotion recognition}

Audio-visual based emotion recognition is to integrate audio and visual modalities with different statistical properties by using fusion strategies at feature, decision and model levels. Feature-level fusion is also called early fusion. A substantial number of previous works~\cite{mansoorizadeh2010multimodal,wang2012kernel} have demonstrated the performances of feature-level fusion on the AVER tasks. However, because it merged audio and visual features in a straightforward way, feature-level fusion could not model the complicated relationships, e.g., the differences on time scales and metric levels, between the two modalities~\cite{zhang2017learning}. Decision-level fusion has also been adopted by almost all the winning systems of the EmotiW challenges~\cite{liu2018multi,li2019bi}. Note that, it is usually implemented by combining the individual classification scores and therefore not able to well capture the mutual correlation among different modalities, as these modalities are assumed to be independent. In~\cite{zeng2008audio}, model-level fusion was performed by fusing audio and visual streams of hidden Markov models (HMMs), which facilitated the building of an optimal connection among multiple streams according to the maximum entropy principle and the maximum mutual information criterion. To improve emotion recognition performances, the mouth area was further divided into several subregions, as elaborated in~\cite{zhao2009lipreading}, to extract LBP-TOP features from each subregion and concatenate the respective features. In~\cite{wang2019multi} a multiple attention fusion network (MAFN) was proposed by modeling human emotion recognition mechanisms.

\section{Proposed Attention and Fusion Strategy}\label{proposed_method}
The proposed multimodal attention and fusion network based on adaptive and multi-level FBP for audio-visual emotion recognition is shown in Fig.~\ref{pipeline}, in which all symbols and numbers will be introduced in the following subsections. Most of audio-visual emotion datasets are annotated at the sentence level~\cite{2006The,busso2008iemocap,dhall2012collecting}. The proposed framework maps a temporal feature sequence to a single label, which mainly consists of three important parts: audio and video encoder, attention, fusion and classifier.

\textbf{Audio and video encoder:} in our study, for audio encoder, a FCN with local response normalization (LRN)~\cite{krizhevsky2012imagenet} is adopted to encode the speech spectrogram into a high-level representation. For video encoder used on AFEW database, each detected face is encoded into a vector through a pre-trained model~\cite{knyazev2018leveraging}, which has been proved to be effective, while on IEMOCAP database, we use the same network as audio stream by utilizing the facial marker information~\cite{2019multimodalbst}.

\textbf{Attention:} 1-D attention-based decoder is employed to obtain information more related to emotion after audio encoder and video encoder respectively.

\textbf{Fusion and classifier:} as for audio-visual fusion, an adaptive and multi-level FBP approach is presented. Based on the fusion vector, the output posterior probabilities of emotion classes can be generated by using a fully-connected (FC) layer followed by a softmax layer. We will elaborate on each module in the following section.

\subsection{Audio stream}\label{propose_audio_system}
\begin{figure}
  \setlength{\belowcaptionskip}{-0.2cm}
  \centering
   \includegraphics[width=40mm]{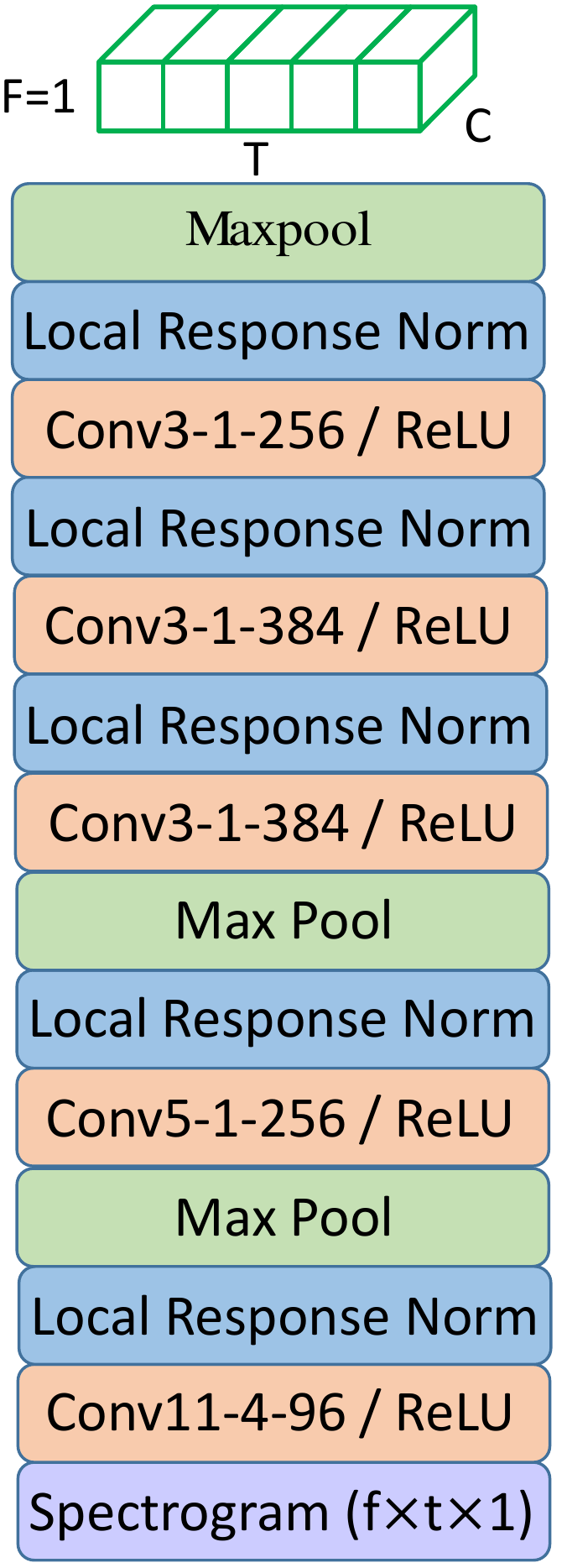}
     \caption{FCN based audio encoder using LRN.}
     \label{fcn_encoder}
\end{figure}

The audio stream directly handles the speech spectrogram by using stacked convolutional layers followed by an attention block. Without handcrafted feature extraction, learning based on CNN has been widely used for SER~\cite{mirsamadi2017automatic,aldeneh2017using}. Inspired by AlexNet~\cite{krizhevsky2012imagenet}, we used a FCN based audio encoder as illustrated in Fig.~\ref{fcn_encoder}. All the convolutional layers are followed by a ReLU activation function and LRN. The dimensions of frequency domain and time domain are f and t respectively.

The main principle of LRN is to suppress neighboring neurons by imitating biological active neurons, aiming at improving the accuracy of deep network training. In~\cite{krizhevsky2012imagenet}, the LRN layer was proposed to create a competition mechanism for the activity of local neurons, which makes the larger response pairs enhanced, and suppresses other neurons with smaller feedback, thus improving the generalization ability of the model. Suppose the activity of a neuron computed by applying kernel $i$ at position $(x,y)$ is denoted by $b_{x,y}^i$. Then by applying the ReLU~\cite{glorot2011deep} nonlinearity, the response-normalized activity $\hat{b}_{x,y}^i$ is expressed as:

\vspace{-0.4cm}

\begin{equation}
\ \hat{b}_{x,y}^i = {{b_{x,y}^i} \mathord{\left/
 {\vphantom {{b_{x,y}^i} {{{\left( {k_L + \alpha \sum\limits_{j = \max (0,i - n/2)}^{\min (N - 1,i + n/2)} {{{(b_{x,y}^i)}^2}} } \right)}^\beta }}}} \right.
 \kern-\nulldelimiterspace} {{{\left( {k_L + \alpha \sum\limits_{j = \max (0,i - n/2)}^{\min (N - 1,i + n/2)} {{{(b_{x,y}^j)}^2}} } \right)}^\beta }}}\
\end{equation}

\noindent where the sum runs over $n$ ¡°adjacent¡± kernel maps at the same spatial position, and $N$ is the total number of kernels in the layer~\cite{krizhevsky2012imagenet}. The constants $k_L$, $n$, $\alpha$, and $\beta$ are hyperparameters. We applied this normalization after applying the ReLU nonlinearity in certain layers.

 To align audio with video in time, here we pool the size of the frequency domain to 1. Accordingly the output of the audio encoder is a 2-D array $ T \times C $ by reshaping, where $T$ and $C$ represent the number of time frames and channels, respectively. We consider the output as a variable-length grid of $T$ elements. Each element is a $C$-dim ($C$=256) vector corresponding to a region of speech spectrogram, represented as ${\boldsymbol a}_i$. Therefore, the whole audio utterance is now denoted as:

\vspace{-0.1cm}

\begin{equation}
\boldsymbol A = \left\{ {{\boldsymbol a}_1, \cdots ,{\boldsymbol a}_T} \right\},{\boldsymbol a}_i \in {\mathbb{R}^C}.\
\end{equation}

\begin{figure}
  \setlength{\belowcaptionskip}{-0.3cm}
  \centering
   \includegraphics[width=65mm]{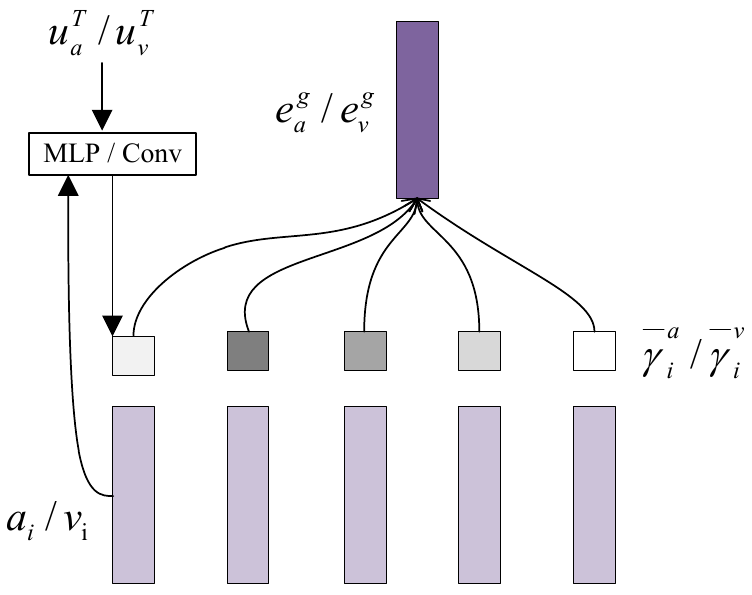}
     \caption{Structure of audio/video self-attention.}
     \label{selfatt}
\end{figure}

Intuitively, not all time-frequency units in set $\boldsymbol A$ contribute equally to the emotion state of the whole utterance. We introduce self-attention to extract the elements that are important to the emotion of the utterance, as shown in Fig.~\ref{selfatt}. By calculating these weights in the time dimension, elements in the set $\boldsymbol A$ are weighted and summed. We use the following formulae to realize this idea:

\vspace{-0.4cm}

\begin{eqnarray}
\gamma_i^a &=& {\boldsymbol{u}_a^\top}\tanh ({\boldsymbol {W}_a}{\boldsymbol {a}_i} + {\boldsymbol {b}_a})\\
\bar{\gamma}_i^a &=& \frac{{\exp ({\lambda _a} {\gamma_i^a})}}{{\sum\nolimits_{k = 1}^T {\exp ({\lambda _a} {\gamma_k^a})} }}\\
{\boldsymbol {e}_a^i} &=& \bar{\gamma} _i^a{\boldsymbol {a}_i}\\
{\boldsymbol {e}_a^g} &=& \sum\limits_{i = 1}^T {\boldsymbol {e}_a^i}
\end{eqnarray}

First, $ {\boldsymbol a}_i $ is fed to a fully connected layer with a parameter set $\{\boldsymbol {W}_a, \boldsymbol {b}_a\}$ followed by a tanh function to obtain a new representation. Then we measure the importance weight $ \gamma_i^a $ by the inner product between the new representation of $ {\boldsymbol a}_i $ and a learnable vector $ \boldsymbol{u}_a $. After that, the normalized importance weight $ \bar{\gamma} _i^a $ is calculated using softmax with the temperature parameter $\lambda _a$ to control the uniformity of the importance weights~\cite{hinton2015distilling}. If $ \lambda _a $ = 0, the weight obtained by attention is the same, which means all the time-frequency units have the same importance to the utterance audio vector ${\boldsymbol {e}_a^g}$. Finally, $ \boldsymbol {e}_a^i $ is computed with importance weights using the set $\boldsymbol A$. By summing $ \boldsymbol {e}_a^i $ over $i$, $ \boldsymbol {e}_a^g $ represents the audio-based global feature vector for emotion.

\subsection{Video stream}\label{propose_video_system}

As shown in Fig.~\ref{pipeline}, for the video stream, the face frames are first detected and aligned using a face detector~\cite{king2009dlib}. Then due to the limited amount of training data in the AVER tasks, such as the EmotiW challenge data set, a pre-trained video encoder network on FER2013~\cite{carrier2013fer} is adopted. The size of the detected faces is $224\times 224$, corresponding to the height and width. If a face is not found on the frame, the entire frame is resampled and passed to the network because the network still could capture some contextual cues and this can also ensure the synchronization of audio and video in time. In~\cite{knyazev2018leveraging}, four kinds of deep convolutional neural networks, e.g., VGG-Face, are adopted to extract emotion-related features from a face image with a 3-channel input (R, G, B). Moreover, three proprietary state-of-the-art face recognition networks, notated as FR-Net-A, FR-Net-B, FR-Net-C, have also been investigated. In this study, FR-Net-B network is chosen as our video encoder as it achieved a high accuracy as demonstrated in~\cite{zhang2019deep,knyazev2018leveraging}.

For each face frame, a $1024$-dim ($D_1$) feature vector of the video encoder output is generated from the last fully connected layer. Therefore the visual feature sequence of an $L$-frame video can be represented as:

\begin{equation}
\ \boldsymbol {V} = \{ {\boldsymbol {v}_1}, \cdots ,{\boldsymbol {v}_L}\} ,{\boldsymbol {v}_i} \in {{\mathbb R}^{D_1}} \
\end{equation}

\noindent where $ \boldsymbol {v}_i $ denotes the facial feature vector of $i$-th frame. Similar to the audio stream, we adopt the self-attention mechanism to calculate the weight for each frame, as shown in Fig.~\ref{selfatt}. Before entering into the attention block, the dimension reduction is conducted to decrease computational complexity and to relieve over-fitting. Here we use a convolution layer as it has fewer parameters than full connection layer. The formulae are listed below:

\begin{eqnarray}
{\gamma_i^v} &=& {\boldsymbol{u}_v^\top}\tanh ({\text{Conv}({\boldsymbol {v}_i})})\\
\bar{\gamma} _i^v &=& \frac{{\exp ({\lambda _v}{\gamma_i^v})}}{{\sum\nolimits_{k = 1}^L {\exp ({\lambda _v}{\gamma_k^v})} }}\\
\boldsymbol {e}_v^i &=& \bar{\gamma} _i^v {{{\text{Conv}({\boldsymbol {v}_i})}}}\\
{\boldsymbol {e}_v^g} &=& \sum\limits_{i = 1}^L {{{\boldsymbol {e}_v^i}}}
\end{eqnarray}
where Conv represents the convolution operation to yield a new low-dimension representation of the $i$-th frame. Specifically, we first extend the input vector ${\boldsymbol {v}_i}$ to a matrix by reshaping. The convolution kernel size is $4\times1$ and the stride is 4 respectively. Then the dimension of Conv output is reduced to $D=256$. Finally, $ \boldsymbol {e}_v^i $ is computed with importance weights using the set $\boldsymbol {V}$. By summing $ \boldsymbol {e}_v^i $ over $i$, $ \boldsymbol {e}_v^g $ represents the video-based global feature vector for emotion.

\subsection{Global factorized bilinear pooling (G-FBP)}\label{G-FBP}
 Bilinear pooling is introduced in~\cite{2015Bilinear} and initially used for feature fusion. Then the fused vectors are used for classification. Although improving the system performance, it also brings a huge amount of computation. Some researches focusing on reducing computational cost have achieved considerable results~\cite{MCBGaoBZD16,li2017factorized}. According to~\cite{yu2017multi}, for the audio feature vector, $ {\boldsymbol {e}_a^g} \in {{\mathbb R}^C} $, and video feature vector, $ {\boldsymbol {e}_v^g} \in {{\mathbb R}^D} $, {the bilinear pooling for the output, $ {I_j} \in {\mathbb R} $, is defined as follows:

\begin{equation}\label{oribp}
\ I_j = {{\boldsymbol {e}_a^g}^\top}{{\boldsymbol{\Lambda}_j}}{\boldsymbol{e}_v^g} \
\end{equation}

\noindent where ${{\boldsymbol{\Lambda}}_j} \in {{\mathbb R}^{C \times D}} $ is the $j$-th projection matrix. By learning a set of projection matrices $ \{ {\boldsymbol \Lambda}_j | j=1,...,O\}$, we can obtain an $O$-dimensional audio-visual fusion vector $ {\boldsymbol {I}} = \left[ {{I_1}, \cdots ,{I_O}} \right] $.

According to~\cite{HadamardLBP,yu2017multi}, the projection matrix $ \boldsymbol{\Lambda}_j $ in Eq.(\ref{oribp}) can be factorized into two low-rank matrices:

\begin{equation}\label{bipool}
\begin{array}{l}
{I_j} = {{\boldsymbol {e}_a^g}^\top}{{\boldsymbol {P}}_j}{\boldsymbol {Q}_j}^\top{\boldsymbol {e}_v^g}\\
\;\;\; = \sum\limits_{d = 1}^K {{{\boldsymbol {e}_a^g}^\top}{{\boldsymbol {p}}_j^d}{{\boldsymbol {q}}_j^d}^\top}{\boldsymbol {e}_v^g} \\
\;\;\; = {{\mathbbm {1}}^\top}({\boldsymbol{P}_j}^\top{\boldsymbol{e}_a^g}\; \circ \;{\boldsymbol{Q}_j}^\top{\boldsymbol{e}_v^g})
\end{array}
\end{equation}

\noindent where $K$ is the latent dimension of the factorized matrices in Eq.(\ref{bipool}), $ {{\boldsymbol {P}}_j} = [{{\boldsymbol {p}}_j^1}, \cdots ,{{\boldsymbol {p}}_j^K}] \in {{\mathbb R}^{C \times K}} $ and $ {{\boldsymbol {Q}}_j} = [{{\boldsymbol {q}}_j^1}, \cdots ,{{\boldsymbol {q}}_j^K}] \in {{\mathbb R}^{D \times K}} $, $ \circ $ represents the element-wise multiplication of two vectors, and $ {\mathbbm{1}} \in {{\mathbb R}^K} $ is an all-1 vector. The advantage is that the low rank matrices ${\boldsymbol{P}}_j$ and ${\boldsymbol{Q}}_j$ are used to approximate $ \boldsymbol{\Lambda}_j $, so the operation is simplified and parameter quantity can be reduced. When we want the output to be a vector, it just need to expand the matrix ${\boldsymbol{P}}_j$ and ${\boldsymbol{Q}}_j$. Specifically, to obtain the output feature vector \textit{${\boldsymbol{I}}$} by Eq.(\ref{facbipool}) below, two 3-D tensors, ${\boldsymbol {P}} = [{{\boldsymbol {P}}_1}, \cdots ,{{\boldsymbol {P}}_O}] \in {{\mathbb R}^{C \times K \times O}} $ and $ {\boldsymbol {Q}} = [{{\boldsymbol {Q}}_1}, \cdots ,{{\boldsymbol {Q}}_O}] \in {{\mathbb R}^{D \times K \times O}} $, need to be learned. Note \textit {$ {\boldsymbol {P}} $} and \textit {$ {\boldsymbol {Q}} $} can be reformulated as 2-D matrices, \textit {${\widetilde {\boldsymbol {P}} \in {{\mathbb R}^{C \times KO}}}$} and \textit {${\widetilde {\boldsymbol {Q}} \in {{\mathbb R}^{D \times KO}}}$}, respectively, by using a reshape operation. Accordingly, we have:

\begin{equation}\label{facbipool}
{\boldsymbol {I}} = {\rm{SumPooling}}({\widetilde {\boldsymbol {P}}^\top}{\boldsymbol {e}_a^g}\; \circ \;{\widetilde {\boldsymbol {Q}}^\top}{\boldsymbol {e}_v^g},K_G)\;
\end{equation}

\noindent where $ {{\widetilde {\boldsymbol {P}}^\top}}{\boldsymbol {e}_a^g} $ and $ {{\widetilde {\boldsymbol {Q}}^\top}}{\boldsymbol {e}_v^g} $ are implemented by feeding \textit {${\boldsymbol {e}_a^g}$} and \textit {${\boldsymbol {e}_v^g}$} to fully connected layers respectively, and the function $ {\rm{SumPooling}}({\boldsymbol {x}},K_G) $ applies sum pooling within a series of non-overlapped windows to \textit {${\boldsymbol {x}}$}. We indicate $\boldsymbol {I}$ in Eq.(\ref{facbipool}) as global factorized bilinear pooling (G-FBP). The G-FBP module is shown in Fig.~\ref{fbpmodule}.

After that, since the magnitude of the output varies dramatically due to the introduced element-wise multiplication, the L2-normalization is used after G-FBP to normalize the energy of \textit {${\boldsymbol {I}}$} to \textit {${\boldsymbol {1}}$}.

\subsection{Adaptive global factorized bilinear pooling (AG-FBP)}

Through our analysis of the experimental results on the EmotiW database (see Section~\ref{exps_analysis} for detail), we find that the influence of audio and video streams on the emotional state of each specific recording is different. For example, it tends to classify audio streams into ``Angry'' and ``Fear'' emotions while video streams are often classified as ``Disgust'', ``Happy'', ``Sad'' or ``Surprise'' emotions. We therefore propose an adaptive strategy for G-FBP (denoted as AG-FBP) audio-visual information fusion. In this study, we adopt the encoder vectors before audio and video fusion to dynamically calculate the two coefficients:

\vspace{-0.2cm}

{
\begin{equation}\label{agfbp_coef1}
\mu = \frac{{\left\| {\boldsymbol {{e}}_a^g} \right\|}}{{\left\| {\boldsymbol {{e}}_a^g} \right\| + \left\| {\boldsymbol {{e}}_v^g} \right\|}}
\end{equation}}

\vspace{-0.3cm}

{\begin{equation}\label{agfbp_coef2}
\eta = \frac{{\left\| {\boldsymbol {{e}}_v^g} \right\|}}{{\left\| {\boldsymbol {{e}}_a^g} \right\| + \left\| {\boldsymbol {{e}}_v^g} \right\|}}
\end{equation}}

\vspace{-0.15cm}

\noindent where $ \mu $ and $ \eta $ are the adaptive factor coefficients of the audio and video streams respectively and computed based on the current sample. $ \left\| \cdot \right\| $ represents the L2-norm operation.

Compared with Eq.(\ref{oribp}), the new formulation is shown below:

\vspace{-0.3cm}

{\begin{equation}
{I_j^\text{A}} = {(\mu {\boldsymbol {e}_a^g})^\top}{{\boldsymbol {\Lambda}}_j}(\eta {\boldsymbol {e}_v^g})
\end{equation}}

\vspace{-0.2cm}

Correspondingly, the formulation of G-FBP in Eq.(\ref{facbipool}) is modified as:
\begin{equation}\label{aptfacbipool}
{\boldsymbol {I}^\text{A}} = {\rm{SumPooling}}({\widetilde {\boldsymbol {P}}^\top}(\mu {\boldsymbol {e}}_a^g)\; \circ \;{\widetilde {\boldsymbol {Q}}^\top}(\eta {\boldsymbol {e}}_v^g),K_G)\;
\end{equation}

In comparison to G-FBP, no additional learning parameters are required in AG-FBP. $ \mu $ and $ \eta $ are adaptively determined by audio-based and video-based global feature vectors for emotion, which are learned in the attention module and can well represent the contribution and correlation of each modality to the current emotion state.

\subsection{Multi-Level factorized bilinear pooling (M-FBP)}
Since the change of emotional state is usually continuous, there is no good characterization of emotional state durations. Some previous studies showed that 250ms was the suggested minimum segment length required for identifying emotion~\cite{provost2013identifying,wollmer2013lstm}. Speech segments have also been investigated for speech emotion recognition~\cite{zhang2017learning,satt2017efficient,chen20183,ChibaNI20}. Emotions change over time, and the audio-visual fusion based on segment level may be more effective. Motivated by this, we proposed a multi-level FBP (M-FBP) approach for audio-visual emotion recognition by using intra-trunk data of one recording. As shown in Section~\ref{G-FBP}, G-FBP only extracts a global audio/video vector ${\boldsymbol {e}}_a^g/{\boldsymbol {e}}_v^g$ for FBP fusion. In addition to G-FBP, M-FBP performs a high-resolution fusion at a segment level, which can fully integrate audio and visual information.

To implement M-FBP, on one hand, the stride of the pooling layer of the audio stream can be modified to adjust the length of intra-trunk audio data, namely $ {\boldsymbol {{e}_a}} = [{\boldsymbol {e}_a^1}, \cdots ,{\boldsymbol {e}_a^H}] $. $H$ is the number of intra-trunks and determined by the time lengths of the sample ($L$) and one intra-trunk ($T$), and $L=H\times T$. On the other hand, the frame rate of the video stream is 40ms while the frame shift of the audio stream is 10ms. To synchronize the audio and video streams, we change the time length of the video stream to be the same as that of the audio stream through the reshape and sum operations, namely ${\boldsymbol{e}_{v}} = [{\boldsymbol {e}}_{v}^1, \cdots ,{\boldsymbol {e}}_{v}^H] $. As the length of each video recording is different, we adopt zero-padding and use masking at the end as in~\cite{zhang2018attention}. Finally, we formulate intra-trunk based FBP as follows:

\vspace{-0.5cm}

\begin{equation}
{\boldsymbol{I}^\text{M}} = \sum\limits_{h = 1}^H {\rm{SumPooling}}\left({{{\widetilde {\boldsymbol{P}}}_h^\top}{{\boldsymbol{e}}_a^h}\; \circ \;{{\widetilde {\boldsymbol{Q}}}_h^\top}{{\boldsymbol{e}}_{v}^h},K_M}\right)
\end{equation}

\noindent where ${\widetilde {\boldsymbol {P}}_h  \in {{\mathbb R}^{C \times KO}}}$ and ${\widetilde {\boldsymbol {Q}}_h  \in {{\mathbb R}^{D \times KO}}}$ are two 3-D tensors for the $h$-th intra-trunk data. Different from G-FBP, the L2-normalization is used after M-FBP to normalize the energy of \textit {${\boldsymbol{I}^\text{M}}$} to \textit {${\boldsymbol {1}}$}.

Note that we can also combine AG-FBP and M-FBP to perform adaptive and multi-level FBP (AM-FBP). Accordingly, both ${\boldsymbol{I}^\text{A}}$ of global-trunk data and ${\boldsymbol{I}^\text{M}}$ of intra-trunk data are concatenated as the fusion vector for the AM-FBP system. For all these audio-visual systems, we update the network parameters using the cross-entropy criterion.

\section{Experiments and Result Analyses}\label{exps_analysis}
To verify the effectiveness of our proposed approach, we validate audio-visual emotion recognition network on IEMOCAP database~\cite{busso2008iemocap} and AFEW8.0 database~\cite{dhall2012collecting} which was used in the audio-visual sub-challenge of EmotiW2019.


The IEMOCAP corpus comprises five sessions, each of which includes labeled emotional speech utterances from recordings of dialogs between two actors. There is no actor overlapping between these sessions. We utilize the database in the same way with~\cite{zhang2018attention}. Four emotional categories are adopted, namely happy, sad, angry and neutral. By only using improvised data instead of acting, we implement a 5-fold cross-validation. The spectrogram extraction process is consistent with~\cite{zhang2018attention}. First, a sequence of overlapping Hamming windows are applied to the speech waveform, with 10 msec window shift, and 40 msec window size. Then, we calculate a discrete Fourier transform (DFT) of length 800 for each frame. Finally, the 200-dimensional low-frequency part of the spectrogram is used as the input. In addition, the IEMOCAP database contains detailed facial marker information from speakers. The Mocap data (facial expression) contains a column of tuples. The sample rate of the marker capture system is 120 frames per second. For details, please refer to~\cite{busso2008iemocap}. The marker point coordinates are used as features for the training of the video-based network. Since we use the facial marker information as the input, 1D-ABFCN which is consistent with audio encoder is also employed as the video encoder. The input channel is 3 and the pool size of the two is different.

The AFEW database is collected from films and TV series to simulate the real world, including seven emotional categories: angry, disgust, fear, happy, neutral, sad, surprise. There are 773 videos and corresponding audios in the training set, 383 in the validation set, and 653 in the test set. We carried out experiments by using the FR-Net-B feature, which has been proved effective in~\cite{zhang2019deep}. For the audio feature, we also used 200-dimensional spectrogram in the same way as IEMOCAP database.

\begin{figure}
 \setlength{\belowcaptionskip}{-0.3cm}
  \centering
   \includegraphics[width=85mm]{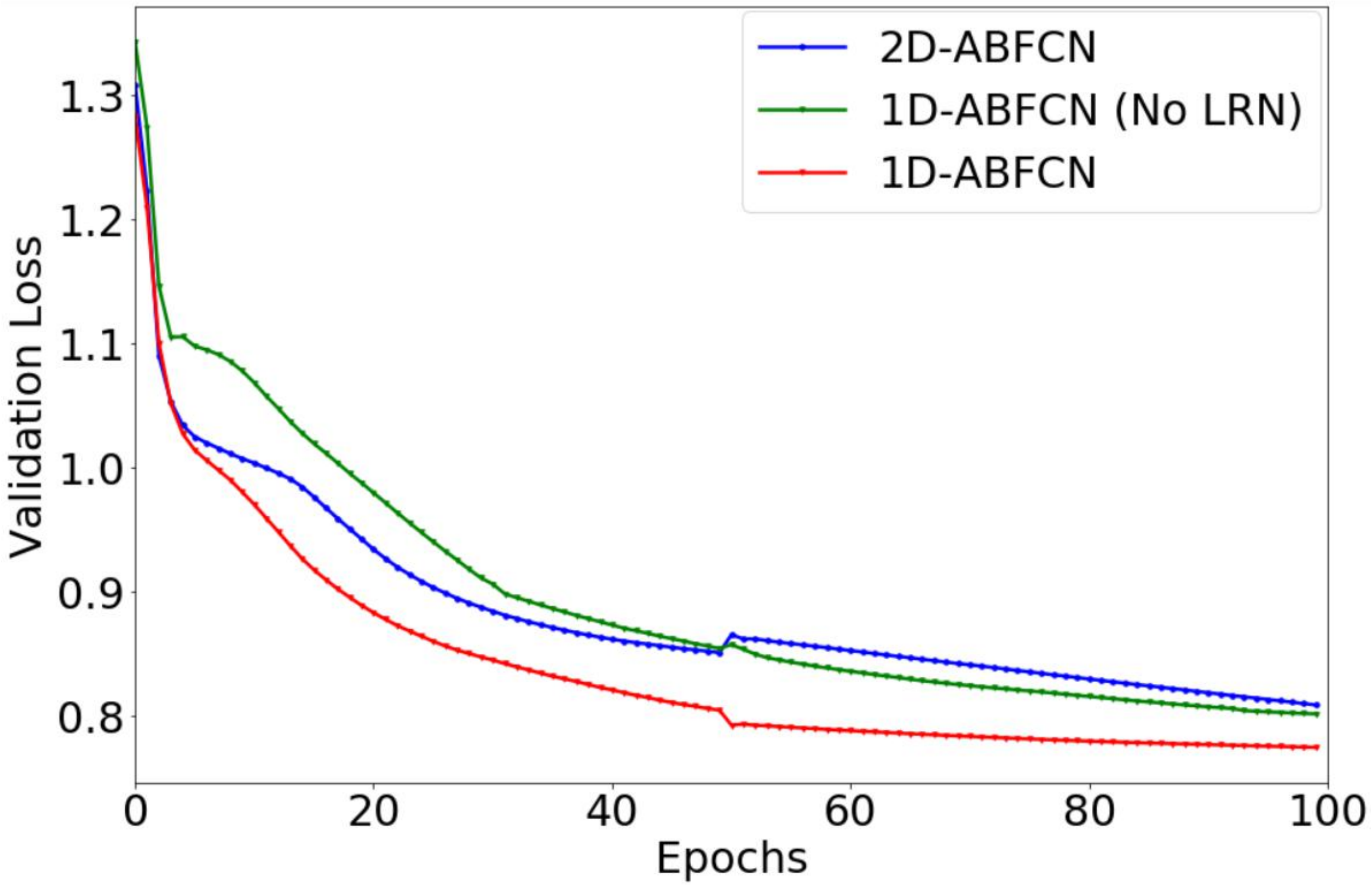}
     \caption{2D-ABFCN and 1D-ABFCN learning curves, where \leftline{ 1D-ABFCN (No LRN) means that only the LRN layer is re-} \leftline{ moved compared with 1D-ABFCN.}}
     \label{lc_2D1D}
\end{figure}

For the audio system and the video system, the proposed architectures are all implemented using TensorFlow 1.2.1 and are trained on two GeForce GTX 1080 GPUs for 100 epochs. The batch size is 32. We use Adam optimizer with $ \lambda _a $ = 0.3, $ \lambda _v $ = 0.5 for the IEMOCAP database, $ \lambda _a $ = 0.3 and $ \lambda _v $ = 1 for the AFEW database. For the audio system, the learning rate is 0.0001. For the video system, the learning rate is 0.0002 for IEMOCAP database and 0.0001 for AFEW database.

For the audio-visual fusion system, the proposed methods are implemented using TensorFlow 1.2.1 and are trained on two GeForce GTX 1080 GPUs for 200 epochs. The batch size is 64. We use Adam optimizer with a learning rate of 0.0001. $ \lambda _a $ = 0.3 and $ \lambda _v $ = 1. We train the whole network with $K_G$ = 4 and $K_M$ = 2. The value of $O$ is 128 in G-FBP system and 192 in M-FBP system. $H = 6$ for the AFEW database and $H = 19$ for the IEMOCAP database. The dropout parameter is set to 0.3 for alleviating the over-fitting problem. The parameters of LRN can refer to~\cite{krizhevsky2012imagenet}.

For the selection of $ \lambda _a $ and $ \lambda _v $ , we carry out experiments with $ \lambda _a $ and $ \lambda _v $ from 0 to 1 with a step of 0.1, and determine the value corresponding to the optimal result on the validation set. And for the selection of $K_G$ and $K_M$, we take 2 as the basic step to train the proposed networks. Finally, $K_G$ and $K_M$ corresponding to the best result on the validation set are determined.

\vspace{-0.2cm}

\subsection{Audio and video based emotion recognition}

\begin{table}[htbp]
\centering
 \setlength{\tabcolsep}{4mm}
 \caption{\label{tab:test1}Classification accuracy comparison of different audio network architectures and parameter initializations on IEMOCAP test set.}
 \begin{tabular}{ccc}
  \toprule
  Systems & Initialization & Accuracy \\
  \midrule
  Att.+BLSTM+FCN~\cite{2019Exploringzhao} & Random & 68.10\% \\
  CNN+LSTM~\cite{satt2017efficient} & Random & 68.80\% \\
  Fusion\_TACN~\cite{LiuLWGGD20} & Random & 69.75\% \\
  2D-ABFCN~\cite{zhang2018attention} & Pre-trained & 70.40\% \\
  1D-ABFCN (No LRN) & Random & 70.79\% \\
  1D-ABFCN & Random & 71.40\% \\
  \bottomrule
 \end{tabular}\label{audio_ser_results}
\end{table}

In order to verify the effectiveness of the proposed audio stream in emotion recognition, we designed our experiments using a 5-fold cross-validation on the IEMOCAP corpus. 1D-ABFCN (No LRN) means that only the LRN layer is removed compared with 1D-ABFCN. The results are listed in Table~\ref{audio_ser_results}. By using the random initialization, our approach yielded an absolute accuracy gain of 2.60\% over the CNN+LSTM based approach. Even in an unfair comparison to 2D-ABFCN with a pre-trained network using ImageNet dataset, the proposed 1D-ABFCN achieved an improvement of 1\% accuracy on the test set. Please note that 1D-ABFCN employed the same configuration and hyperparameter setting as 2D-ABFCN. The effectiveness of this initialization method based on ImageNet dataset on speech emotion recognition has been investigated in~\cite{zhang2018attention}, and that really gets a higher accuracy than random initialization. Compared with the Fusion\_TACN~\cite{LiuLWGGD20}, the performance of our proposed method is also improved by 1.65\%. After using LRN layer, the system accuracy is increased by 0.61\%. Furthermore, we make a comparison of the learning curves between 2D-ABFCN, 1D-ABFCN (No LRN) and 1D-ABFCN on the validation set in Fig.~\ref{lc_2D1D}. After 50 epochs, the learning rate was halved. By using LRN in 1D-ABFCN, the generalization ability of the model can be enhanced, and the convergence can be accelerated. Accordingly, our approach can achieve a smaller loss which leads to a higher accuracy.

\begin{figure*}[htbp]
\centering

\subfigure[]{
\centering
\includegraphics[width=3.15in]{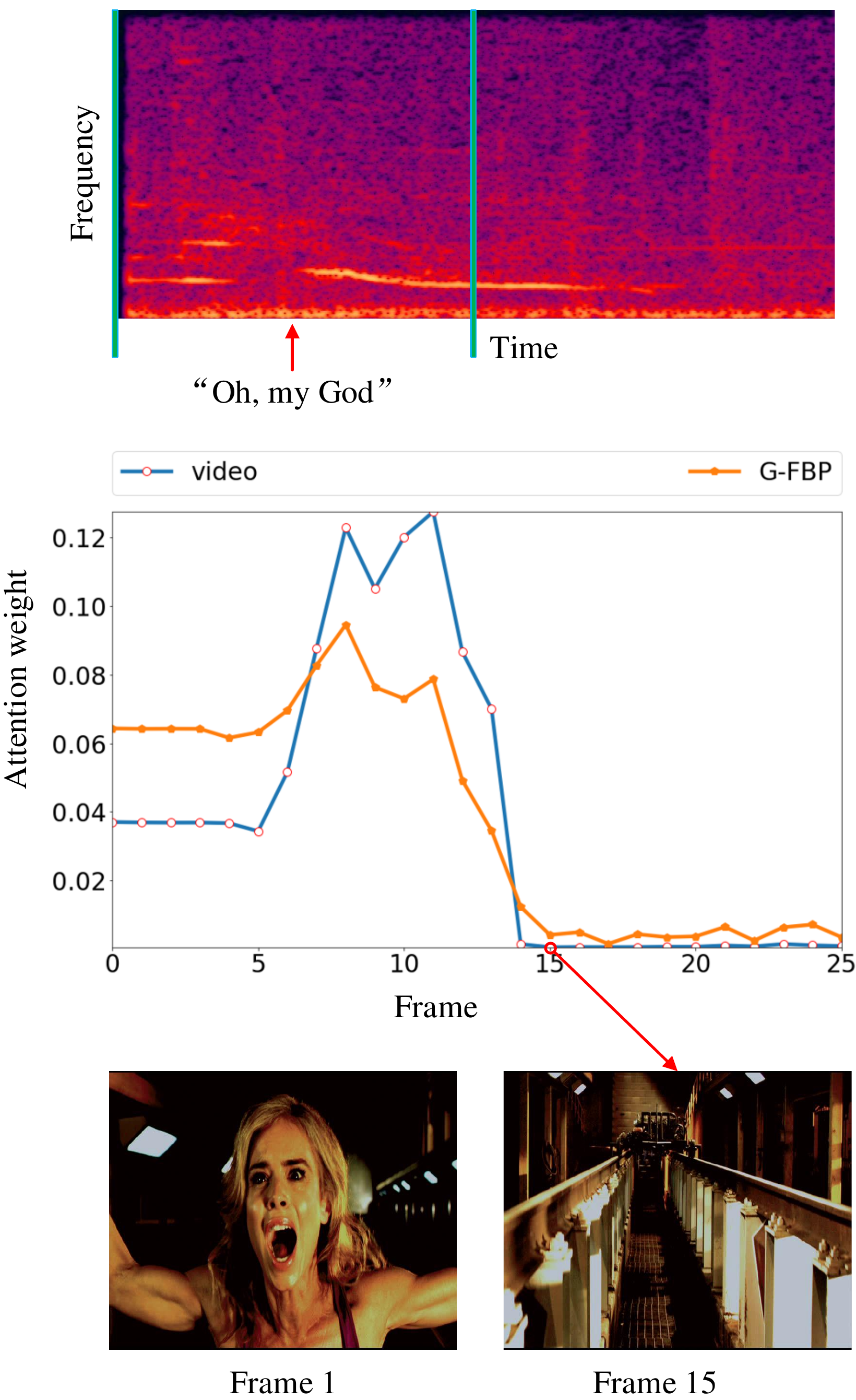}
}
\subfigure[]{
\centering
\includegraphics[width=3.25in]{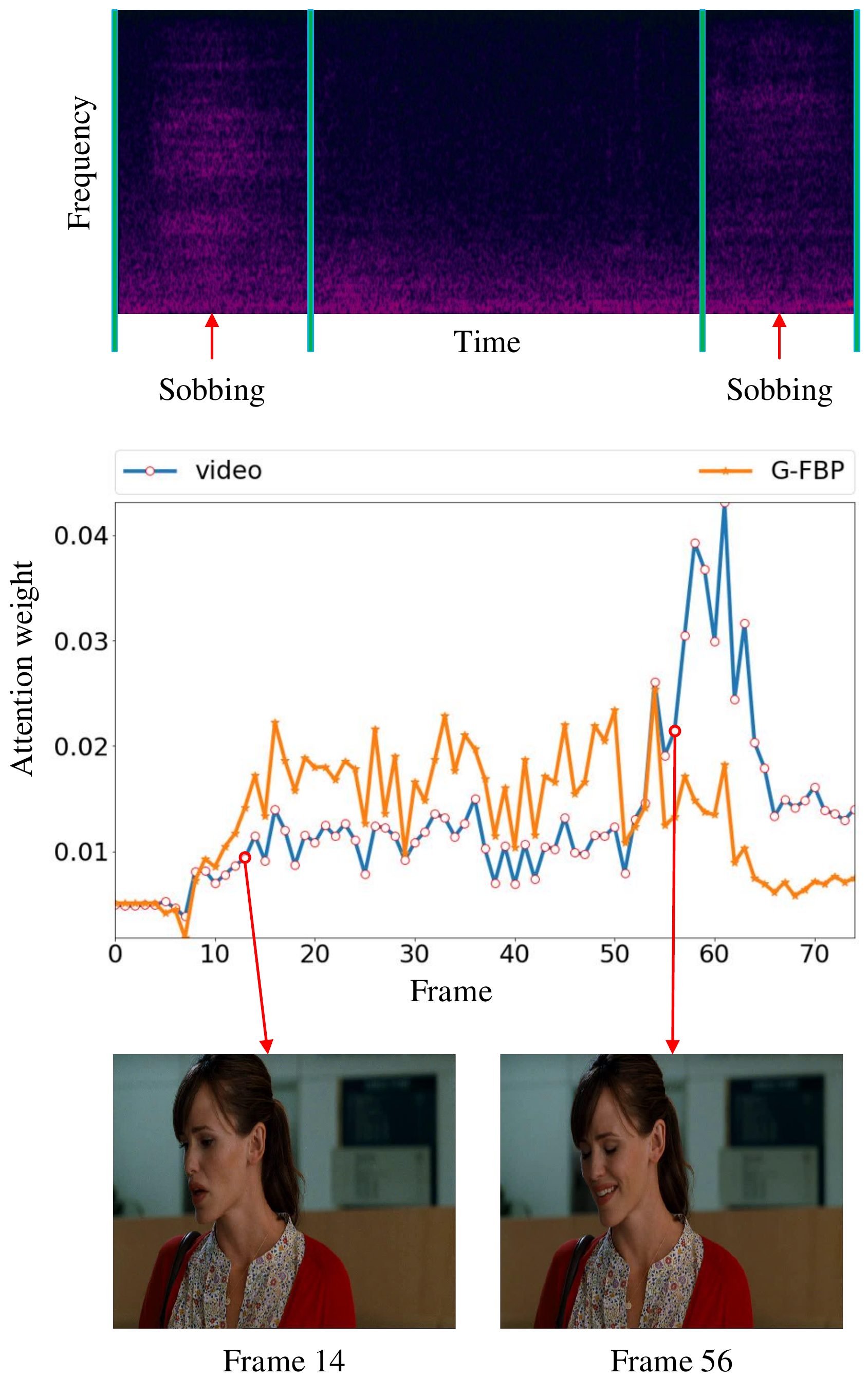}
}%
\centering
\caption{Attention analysis of two randomly selected examples. For each, the curves show the video attention weights along frames in the video and the audio-visual G-FBP systems, with corresponding audio spectrograms and partial video frames.}
\label{AV_attention}
\end{figure*}

\begin{table}[htbp]
\centering
 \setlength{\tabcolsep}{6mm}
 \caption{\label{tab:test2}Classification accuracy comparison of different systems on AFEW validation set.}
 \begin{tabular}{ccc}
  \toprule
  Systems & Accuracy \\
  \midrule
 EmotiW2019 baseline~\cite{dhall2019emotiw}  & 38.81\% \\
 Audio system &  34.99\% \\
 Video system &  52.07\% \\
 G-FBP &  61.10\% \\
  \bottomrule
 \end{tabular}\label{gfbp_results}
\end{table}

\vspace{-0.3cm}

\subsection{Audio-visual emotion recognition with G-FBP}

On top of the 1D-ABFCN based audio stream, we next evaluate audio-visual fusion using G-FBP. We show in Table~\ref{gfbp_results} accuracies obtained with different systems on the AFEW validation set. Compared with the EmotiW2019 baseline audio-visual system~\cite{dhall2019emotiw}, our proposed G-FBP approach significantly improved the accuracy from 38.81\% to 61.10\%. As an ablation study of different modalities, our audio system in Section~\ref{propose_audio_system} and video system in Section~\ref{propose_audio_system} yield the emotion recognition accuracy of 34.99\% and 52.07\%, respectively. Clearly, G-FBP fusion is quite effective with an accuracy gain of about 9\% over the single video modality.

To better demonstrate the effect of G-FBP fusion via the attention mechanism, we plot the attention weights of each frame for two randomly picked examples from the validation set in Fig.~\ref{AV_attention}. For each curve, the values represent the video attention importance of each frame, the higher the better, in the video-only and audio-visual G-FBP systems.

For the example in Fig.~\ref{AV_attention}(a), we show that the video attention can well align with the audio and video frames. After 14 frames, no face was detected in the video, where the video frame could be completely regarded as noise for emotion recognition. Accordingly, the corresponding attention weights in the video system were all zeros. However, in the audio-visual fusion system, considering that the first few audio frames contained women's voice of fear and the last few audio frames contained machine noise, the weight of the video stream changed accordingly. Specifically, the weight of the first seven frames increases due to the enhanced emotion with the audio modality while the weight after 14 frames slightly fluctuated due to the machine noise, which demonstrated the good coupling between audio and video modalities. The video system classified this example as ``Surprise'' by inputting facial features, and it was correctly classified as ``Fear'' after audio-visual fusion.

For the example in Fig.~\ref{AV_attention}(b), we illustrate the complementarity between audio and video modalities using the video attention weight. In the first few frames, the woman was facing sideways, and gradually turned to the front. The corresponding attention weight was also increasing in the video system. After about 56 frames, the woman smiled bitterly, leading to much larger weights. The video system classified this example as ``Disgust'' while the ground truth is ``Sad''. This classification error might be generated due to the wry smile of the woman. Moreover, this example was also misclassified as ``Neutral'' by the audio system. This could be explained by the spectrogram that only starting and ending periods included the weak emotion information of a little sob. However, by the influence of the audio stream, the attention weight of the video stream in the G-FBP fusion system slowly paid attention to the 14-62 frames with a slightly sad face. At the same time, it also reduced the weights during the period of a wry smile. Although both audio system and video system generated wrong emotion classes by partial or weak emotion information, correct classification could be still achieved through the G-FBP system by deep fusion of both audio and video information.

\begin{table}
\centering
\setlength{\tabcolsep}{0.90mm}
\caption{The accuracy on the AFEW validation set for single-modality emotion classification.}
\linespread{1.0}\selectfont
\begin{tabular}{|l|l|l|l|l|l|l|l|}
\hline
\makecell[c]{Modality} & \makecell[c]{Angry} & \makecell[c]{Disgust} & \ \makecell[c]{Fear} \ & \makecell[c]{Happy} & \ \ \makecell[c]{Sad} \ \ & \makecell[c]{Surprise} & \makecell[c]{Neutral} \\ \hline
\makecell[c]{Audio} & \makecell[c]{81.25\%} & \makecell[c]{0.00\%} & \makecell[c]{34.78\%} & \makecell[c]{14.29\%} & \makecell[c]{21.31\%} & \makecell[c]{0.00\%} & \makecell[c]{66.67\%} \\ \hline
\makecell[c]{Video} & \makecell[c]{59.38\%} & \makecell[c]{25.00\%} & \makecell[c]{10.87\%} & \makecell[c]{80.95\%} & \makecell[c]{55.74\%} & \makecell[c]{39.13\%} & \makecell[c]{60.32\%} \\ \hline
\end{tabular}\label{agfbp_analysis}
\end{table}

\begin{figure}
  \centering
   \includegraphics[width=88mm]{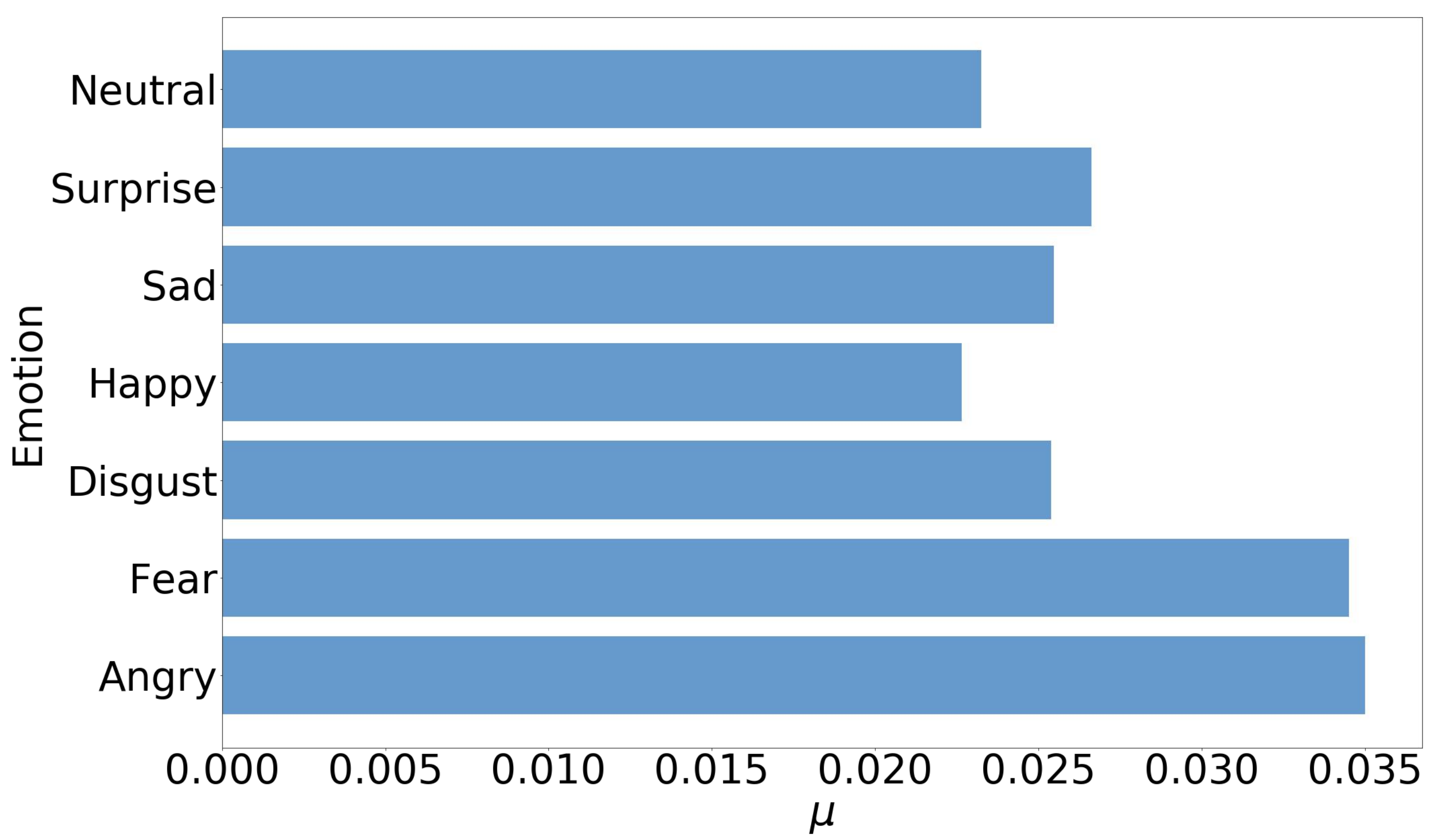}
     \caption{Adaption factors of the audio modality in AG-FBP.}
     \label{agfbp_factor}
\end{figure}

\subsection{Audio-visual emotion recognition with AG-FBP}

As listed in Table~\ref{agfbp_analysis}, we analyze the single-modality classification results of audio and video systems on the AFEW validation set. The numbers in the table represent the accuracy for each emotion. We observed that the impact of audio and video on each emotional category was quite different. Specifically, although the overall accuracy of audio-only system is much lower than that of video-only system (see Table~\ref{gfbp_results}), the accuracy of ``Angry'' and ``Fear'' for the audio modality is significantly higher than that of video system, which shows that for these two emotions, audio modality usually played a more important role. While in terms of ``Happy'', ``Sad'', ``Disgust'' and ``Surprise'', video modality information seemed more discriminative, especially for the ``Happy'' category. This is the main motivation to propose AG-FBP by adaptively determining the importance of each modality for one specific sample or recording.

From the results on the AFEW validation set in Table~\ref{fbps_validation},we achieved an absolute accuracy increase of 1.30\% from G-FBP to AG-FBP. It is worth noting that AG-FBP does not yield additional overhead in terms of storage and computation over G-FBP as only an adaption factor in Eq.(\ref{agfbp_coef1}) needs to be calculated for a specific recording. By visualizing the mean of adaption factors for each emotion category in Fig.~\ref{agfbp_factor}, we observed that the adaptive weighting factors and the relative strength of the modality for each category in Table~\ref{agfbp_analysis} were well consistent. For example, the adaption factors of ``Fear'' and ``Angry'' in the audio modality are higher than those of the other categories, while for ``Happy'', it is the lowest. These demonstrate that, for recordings of different emotions, the representation ability of the audio and video modalities can be quite different.

\begin{table}
\centering
 \setlength{\tabcolsep}{1mm}
 \caption{\label{tab:test3}Classification accuracy and p-value of our improved FBP approaches on the AFEW validation set.}
 \begin{tabular}{ccc}
  \toprule
  Systems &  Accuracy & p-value \\
  \midrule
  G-FBP &  61.10\% & - \\
  AG-FBP & 62.40\% & 0.004 \\
  M-FBP & 63.18\% & 0.002\\
  AM-FBP & 64.17\% & 0.001\\
  \bottomrule
 \end{tabular}\label{fbps_validation}
\end{table}

\begin{figure}
  \centering
    \includegraphics[width=80mm]{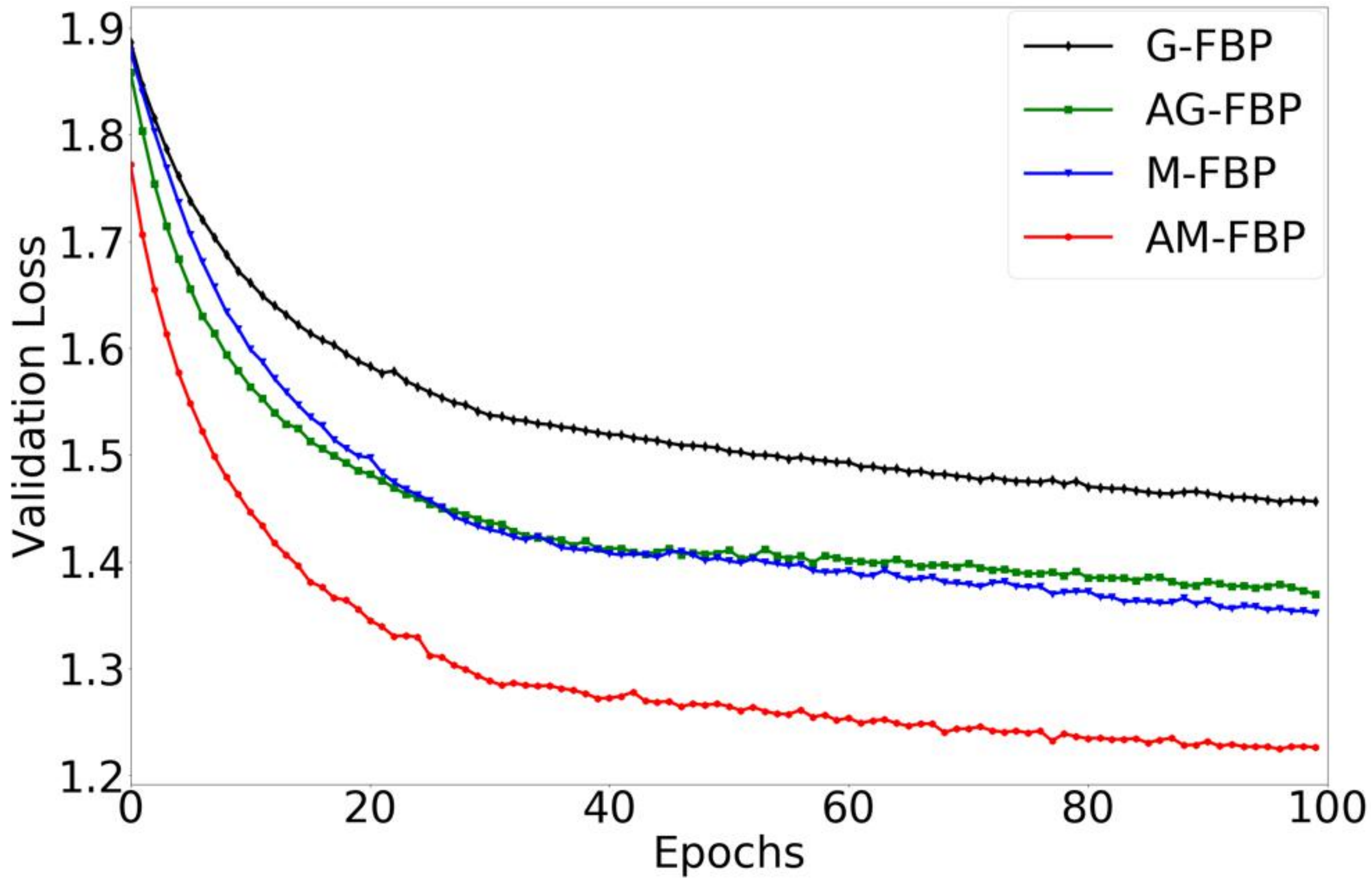}
     \caption{Learning curves of different FBP systems on the validation set.}
     \label{lcurve_fbp}
\end{figure}

\begin{figure*}
  \centering
   \includegraphics[width=160mm]{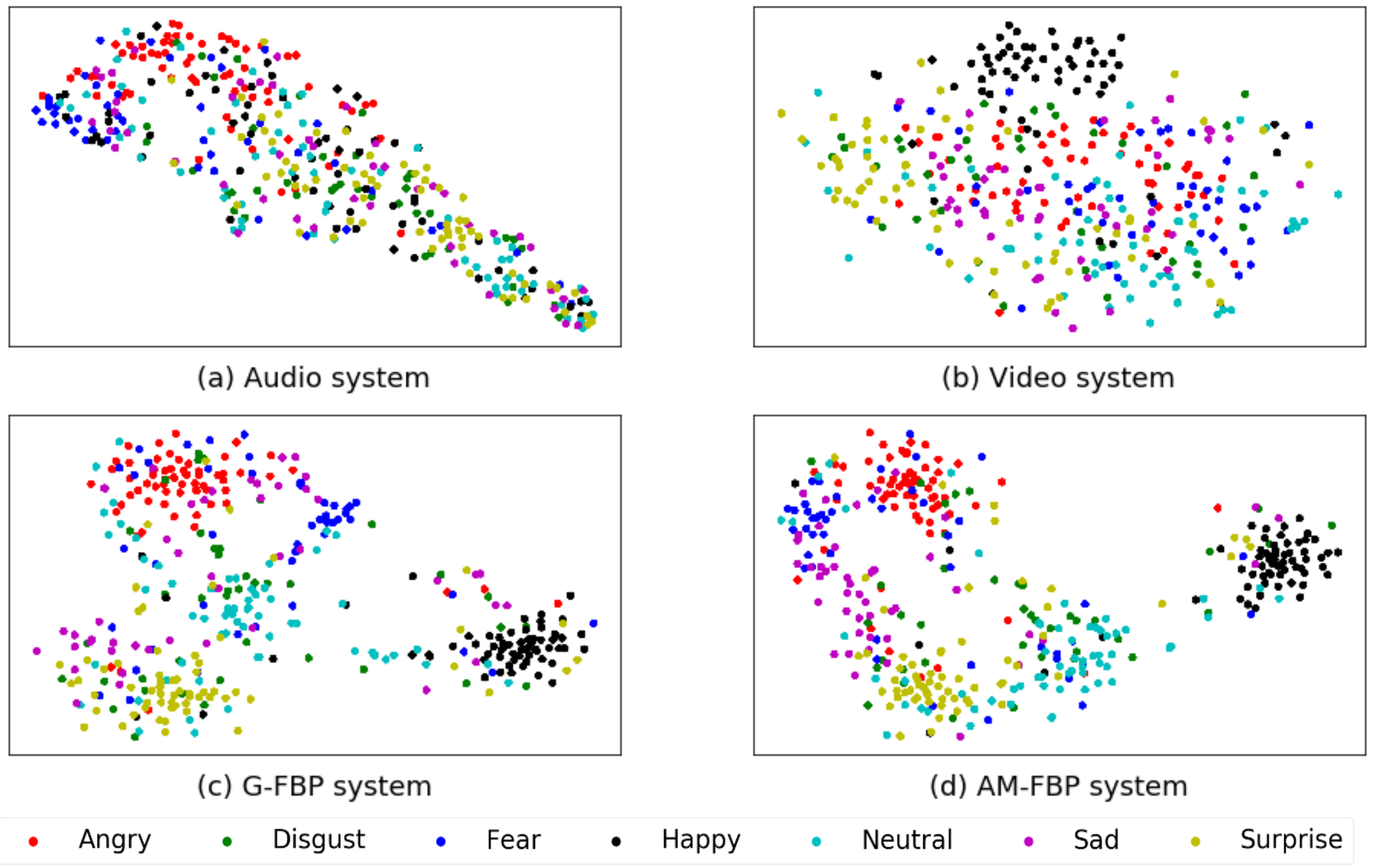}
     \caption{Embedding visualization of different network architectures for emotion recognition.}
     \label{embbeding_vis}
\end{figure*}

\subsection{Audio-visual emotion recognition with M-FBP/AM-FBP}

Next, the effects of multi-level FBP and its combination with AG-FBP are discussed. As shown in Table~\ref{fbps_validation}, we see the M-FBP approach achieves the accuracy of 63.18\%, yielding a gain of 0.78\% over the AG-FBP system by using the multi-level information of both global-trunk and intra-trunk data. By fully utilizing the complementarity between AG-FBP and M-FBP, the proposed AM-FBP system attains the best accuracy of 64.17\% among all the improved FBP approaches. We perform the significance test which is a one-tailed test with the null hypothesis that there is no performance difference between the two systems. Here we refer to the ``Matched Pair Test'' method mentioned in~\cite{115546} to calculate the p-values, which are the main indicator of significance tests, and reflect the degree of support for the null hypothesis. The smaller p-value is, the bigger significant differences between two systems are. The p-values between the G-FBP and our improved FBP systems on the AFEW validation set are listed in Table~\ref{fbps_validation}, which imply that there is a high probability that our improved FBP approaches are able to achieve a better performance compared to the G-FBP approach.

As illustrated in Fig.~\ref{lcurve_fbp}, we compare the learning curves of four audio-visual fusion strategies using cross entropy on the validation set. First of all, the proposed AG-FBP framework shows a lower learning loss and attains a better accuracy than the G-FBP system. As expected the adaptive M-FBP framework also learns well and achieve more stable and better convergence behaviors than the M-FBP system when comparing the bottom curve with the M-FBP curve above it in Fig.~\ref{lcurve_fbp}, which is also consistent with the recognition results in Table~\ref{fbps_validation}.

In Fig.~\ref{embbeding_vis}, we visualize the embedding in the fully-connected layer of different network architectures for emotion recognition on the validation set using t-SNE~\cite{maaten2008visualizing}. Clearly, for single-modality audio or video system, the overall embedding seemed scattered in general for different emotion categories. On the other hand, ``Angry'' was more clustered in the audio system while ``Happy'' was easier to distinguish in the video system. However, in the multi-modal systems of G-FBP and AM-FBP, the embedding was much more distinctive than that of single-modality systems. For example, the G-FBP system could well distinguish both the ``Angry'' and ``Happy'' categories. When compared with G-FBP, AM-FBP yielded the best embedding results with clearer boundaries among different colors (categories), demonstrating its effectiveness in deeper interactions between audio and video modalities. All those results are well aligned with the classification accuracies.

\subsection{Overall comparison}

\begin{table}
\centering
 \setlength{\tabcolsep}{1mm}
 \caption{The overall performance comparison and p-value of different systems on the AFEW test set.}
 \begin{tabular}{cccc}
  \toprule
  Systems & Single model & Accuracy & p-value \\
  \midrule
  EmotiW2019 baseline~\cite{dhall2019emotiw} & ${\checkmark}$ & 41.07\% & - \\
  MAFN~\cite{wang2019multi}  & $\times $  & 58.65\%  & - \\
  4CNNs+LMED+DL-A+LSTM~\cite{liu2018multi} & $\times $  & 61.87\%  & - \\
  4CNNs+BLSTM+Audio~\cite{li2019bi} & $\times $  & 62.78\%  & - \\
  \midrule
  G-FBP  & ${\checkmark}$  & 60.64\%   & - \\
  4G-FBP  & $\times$  & 62.48\%  & - \\
  AG-FBP  & ${\checkmark}$  & 61.26\%  & 0.008 \\
  M-FBP & ${\checkmark}$  & 61.87\%  & 0.001 \\
  AM-FBP  & ${\checkmark}$  & 62.17\%  & \textless 0.001 \\
  2AM-FBP  & $\times$  & 62.79\%  & \textless 0.001 \\
  2AM-FBP+4G-FBP & $\times $  & \bfseries{63.09}\%  & \textless 0.001 \\
  \bottomrule
 \end{tabular}\label{overall_comp}
\end{table}

\begin{figure}
  \centering
   \includegraphics[width=85mm]{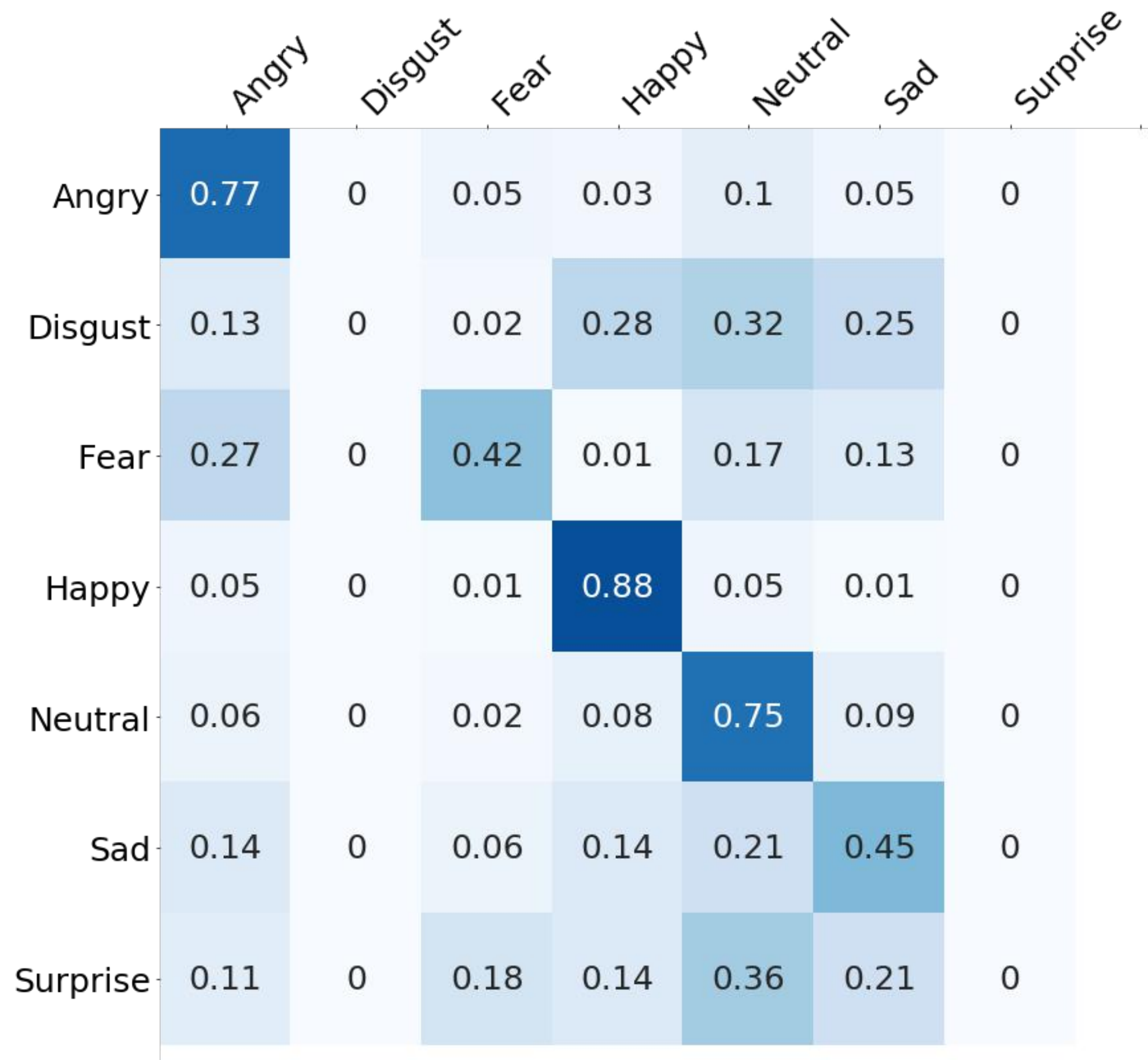}
     \caption{Confusion matrix on AFEW test set.}
     \label{cm_best_test}
\end{figure}

To perform an overall comparison among our proposed techniques on the EmotiW2019 challenge data, we add the validation data to the training set similar to other participating teams in EmotiW~\cite{hu2017learning,fan2016video}. In the upper block of Table~\ref{overall_comp}, we show the performance of different systems on the AFEW test set. MAFN~\cite{wang2019multi} is a multi-modal adaption method with intra/inter-modality attention mechanisms. ``4CNNs+LMED+DL-A+LST''~\cite{liu2018multi} combined five visual and two audio models and obtained the best accuracy in EmotiW2018. While ``4CNNs+BLSTM+Audio''~\cite{li2019bi} fused four visual and three audio models to rank as the champion system for the EmotiW2019 challenges. It also obtained the best accuracy of 62.78\% as shown in Table~\ref{overall_comp}. In the bottom block, using the same test set to evaluate our four proposed FBP systems, our proposed AM-FBP system with a single model setting can yield a competitive accuracy of 62.17\%, indicating the effectiveness of AM-FBP fusion. Moreover, the results of G-FBP and M-FBP on the test set also followed a similar accuracy trend to those on the validation set. In order to ensure the generalization ability of the whole network, we randomly selected at least two groups of models trained with different random seeds for proposed framework to test, e.g., ``4G-FBP'' and ``2AM-FBP'' shown in Table~\ref{overall_comp}. And it can be observed that the results after fusion are also improved. Finally, we integrated two AM-FBP and four G-FBP models to achieve the best result of 63.09\% among all systems as shown in the bottom row in Table~\ref{overall_comp}. We further perform the significance analysis. The p-values between the G-FBP and our improved FBP systems on the AFEW test set are listed in Table~\ref{overall_comp}, which imply that there is a high probability that our improved FBP approaches are able to achieve a better performance compared to the G-FBP approach.

For a further analysis in Fig.~\ref{cm_best_test}, we illustrate the confusion matrix of our best AM-FBP system on the test set. We can observe that ``Angry'', ``Happy'' and ``Neutral'' were more easily classified. As for ``Surprise'' and ``Disgust'', the worse performance might be due to a potential mixing of different emotions, making these emotion categories not easy to be correctly classified. We also observe that the proportion of these two emotions is the lowest in the training set, and similar results are also found in~\cite{li2019bi,liu2018multi,wang2019multi,3264992Wang}.

\begin{table}[htbp]
\centering
 \setlength{\tabcolsep}{6mm}
 \caption{\label{tab:test4}Classification accuracy comparison and p-value of different systems on IEMOCAP test set.}
 \begin{tabular}{ccc}
  \toprule
  Systems & Accuracy & p-value \\
  \midrule
  Audio system~\cite{Zhengem2021} & 63.00\% & - \\
  Decision fusion~\cite{Zhengem2021} & 65.40\% & - \\
  Audio system~\cite{2019multimodalbst} & 50.97\% & - \\
  Video system~\cite{2019multimodalbst} & 49.39\% & - \\
  Encoder concat~\cite{2019multimodalbst} & 67.58\% & - \\
 \bottomrule
  Audio system & 71.40\% & - \\
  Video system & 53.42\% & - \\
  Decision fusion & 72.54\% & - \\
  Encoder concat & 73.11\% & - \\
  G-FBP & 73.98\% & 0.001 \\
  AM-FBP & 75.49\% & \textless 0.001 \\
 \bottomrule
 \end{tabular}\label{iemocap audiovisual_results}
\end{table}

Finally, the proposed methods are further evaluated on the IEMOCAP database. We implemented video-based emotion recognition by using the 1D-ABFCN network and the results are shown in Table~\ref{iemocap audiovisual_results}. In the upper block of Table~\ref{iemocap audiovisual_results}, we show the performance of different systems on the IEMOCAP test set. Different machine learning and deep learning based models are designed to investigate multimodal emotion recognition in~\cite{2019multimodalbst}. In~\cite{Zhengem2021}, feature extraction scheme and matching model structure are investigated for multimodal emotion recognition respectively. Although it does not show the performance of the video-based system in~\cite{Zhengem2021}, the audio-visual system improves by 2.40\% compared with the audio-based system through decision fusion. In the bottom block, inspired by~\cite{2019multimodalbst} and~\cite{Zhengem2021}, the two common fusion methods, namely ``Encoder concat'' and ``Decision fusion'', are selected to compare with the proposed AM-FBP approach. According to Table~\ref{iemocap audiovisual_results}, the accuracy of encoder concat is 0.57\% higher than that of decision fusion and we further take the encoder concat as the audio-visual fusion baseline. From Table~\ref{iemocap audiovisual_results}, we can observe that the proposed AM-FBP achieves the highest accuracy of 75.49\% with the absolute accuracy gain of 2.38\% compared with encoder concat.

It is worth noting that for the single modality systems, the accuracy of the video-based system is higher on the AFEW database than that of audio-based system, while the observation is opposite on the IEMOCAP database. According to the results of the audio-visual system, the two modalities are more complementary on the AFEW database. This result may originate from the differences of the two databases. First, these two databases were collected in different ways. The AFEW database was collected from films and TV series with very complex scenarios and environments, which is a great challenge to the audio-based system. While the IEMOCAP database was recorded from only ten actors in the lab with defined scenes, and we believe it is easier for the audio-based system. Second, the facial features in the two databases vary in the level of detail. For the AFEW database, the detected faces are sent to the trained neural network to extract high-level embedded features. While for the IEMOCAP database, the publisher provides detailed information about the actor's facial expression by placing the markers on the actor's face, head, and hands, which is also widely utilized, e.g., in~\cite{2019multimodalbst} and~\cite{Zhengem2021}. Due to the sparse markers, the emotional cues of video are also limited. From the above two perspectives, it is more difficult to correctly recognize emotions in the AFEW database than in the IEMOCAP database, for example, one obvious observation is that the audio-based system can achieve better performance on the IEMOCAP database (71.40\%) than the AFEW database (34.99\%). Despite these differences, the performances of audio-visual fusion by using the proposed AM-FBP are improved on both two databases, which demonstrates a good generalization ability of the proposed method. The p-values between our FBP systems and the encoder concat on the test set are also shown in Table~\ref{iemocap audiovisual_results}, which imply that the superiority of the proposed FBP is statistically significant.

\section{Conclusion}\label{conclusion}
In this paper, we introduce a novel audio-visual emotion recognition attention network using adaptive and multi-level FBP fusion. Specifically, the deep features from the audio encoder and the video encoder are first selected through the embedding attention mechanism to obtain the emotion-related regions for FBP fusion. Then, the adaptive adjustment of audio and video weights is presented for a deep fusion. Furthermore, the multi-level information by using global-trunk data and intra-trunk data is adopted to design a new network architecture. The proposed approach is verified on the test set of the EmotiW2019 challenge and IEMOCAP database, outperforming other state-of-the-art approaches in literature.

\section*{Acknowledgment}
The authors would like to thank the organizers of EmotiW for evaluating the accuracies of the proposed systems on the test set of the AFEW database. This work was supported by the Strategic Priority Research Program of Chinese Academy of Sciences under Grant No. XDC08050200.



\bibliographystyle{IEEEtran}
\bibliography{myreference}

\begin{thebibliography}{10}
\providecommand{\url}[1]{#1}
\csname url@samestyle\endcsname
\providecommand{\newblock}{\relax}
\providecommand{\bibinfo}[2]{#2}
\providecommand{\BIBentrySTDinterwordspacing}{\spaceskip=0pt\relax}
\providecommand{\BIBentryALTinterwordstretchfactor}{4}
\providecommand{\BIBentryALTinterwordspacing}{\spaceskip=\fontdimen2\font plus
\BIBentryALTinterwordstretchfactor\fontdimen3\font minus
  \fontdimen4\font\relax}
\providecommand{\BIBforeignlanguage}[2]{{%
\expandafter\ifx\csname l@#1\endcsname\relax
\typeout{** WARNING: IEEEtran.bst: No hyphenation pattern has been}%
\typeout{** loaded for the language `#1'. Using the pattern for}%
\typeout{** the default language instead.}%
\else
\language=\csname l@#1\endcsname
\fi
#2}}
\providecommand{\BIBdecl}{\relax}
\BIBdecl

\bibitem{ren2019multi-modal}
M.~Ren, W.~Nie, A.~Liu, and Y.~Su, ``Multi-modal correlated network for emotion
  recognition in speech,'' \emph{Visual Informatics}, vol.~3, no.~3, pp.
  150--155, 2019.

\bibitem{bowers2006faces}
D.~Bowers, K.~Miller, W.~Bosch, D.~Gokcay, O.~Pedraza, U.~Springer, and M.~S.
  Okun, ``Faces of emotion in parkinsons disease: Micro-expressivity and
  bradykinesia during voluntary facial expressions,'' \emph{Journal of The
  International Neuropsychological Society}, vol.~12, no.~6, pp. 765--773,
  2006.

\bibitem{yuvaraj2014detection}
R.~Yuvaraj, M.~Murugappan, N.~M. Ibrahim, K.~Sundaraj, M.~I. Omar, K.~Mohamad,
  and R.~Palaniappan, ``Detection of emotions in parkinson's disease using
  higher order spectral features from brain's electrical activity,''
  \emph{Biomedical Signal Processing and Control}, vol.~14, pp. 108--116, 2014.

\bibitem{vinola2015a}
C.~Vinola and K.~Vimaladevi, ``A survey on human emotion recognition
  approaches, databases and applications,'' \emph{Electronic Letters on
  Computer Vision and Image Analysis}, vol.~14, no.~2, pp. 24--44, 2015.

\bibitem{satt2017efficient}
A.~Satt, S.~Rozenberg, and R.~Hoory, ``Efficient emotion recognition from
  speech using deep learning on spectrograms.'' in \emph{18th Annual Conference
  of the International Speech Communication Association, INTERSPEECH}, 2017,
  pp. 1089--1093.

\bibitem{dhall2018emotiw}
A.~Dhall, A.~Kaur, R.~Goecke, and T.~Gedeon, ``Emotiw 2018: Audio-video,
  student engagement and group-level affect prediction,'' in \emph{2018
  International Conference on Multimodal Interaction (ICMI)}, 2018, pp.
  653--656.

\bibitem{dhall2019emotiw}
A.~Dhall, ``Emotiw 2019: Automatic emotion, engagement and cohesion prediction
  tasks,'' in \emph{2019 International Conference on Multimodal Interaction
  (ICMI)}, 2019, pp. 546--550.

\bibitem{gunes2007bi}
H.~Gunes and M.~Piccardi, ``Bi-modal emotion recognition from expressive face
  and body gestures,'' \emph{Journal of Network and Computer Applications},
  vol.~30, no.~4, pp. 1334--1345, 2007.

\bibitem{bong2017implementation}
S.~Z. Bong, K.~Wan, M.~Murugappan, N.~M. Ibrahim, Y.~Rajamanickam, and
  K.~Mohamad, ``Implementation of wavelet packet transform and non linear
  analysis for emotion classification in stroke patient using brain signals,''
  \emph{Biomedical signal processing and control}, vol.~36, pp. 102--112, 2017.

\bibitem{tian2001recognizing}
Y.-I. Tian, T.~Kanade, and J.~F. Cohn, ``Recognizing action units for facial
  expression analysis,'' \emph{IEEE Transactions on pattern analysis and
  machine intelligence}, vol.~23, no.~2, pp. 97--115, 2001.

\bibitem{chen20183}
M.~Chen, X.~He, J.~Yang, and H.~Zhang, ``3-d convolutional recurrent neural
  networks with attention model for speech emotion recognition,'' \emph{IEEE
  Signal Processing Letters}, vol.~25, no.~10, pp. 1440--1444, 2018.

\bibitem{zhang2018attention}
Y.~Zhang, J.~Du, Z.~Wang, J.~Zhang, and Y.~Tu, ``Attention based fully
  convolutional network for speech emotion recognition,'' in \emph{2018
  Asia-Pacific Signal and Information Processing Association Annual Summit and
  Conference (APSIPA ASC)}.\hskip 1em plus 0.5em minus 0.4em\relax IEEE, 2018,
  pp. 1771--1775.

\bibitem{8470342}
Y.~Li, J.~Tao, B.~Schuller, S.~Shan, D.~Jiang, and J.~Jia, ``Mec 2017:
  Multimodal emotion recognition challenge,'' in \emph{2018 First Asian
  Conference on Affective Computing and Intelligent Interaction (ACII
  Asia)}.\hskip 1em plus 0.5em minus 0.4em\relax IEEE, 2018, pp. 1--5.

\bibitem{li2018attention}
P.~Li, Y.~Song, I.~Mcloughlin, W.~Guo, and L.~Dai, ``An attention pooling based
  representation learning method for speech emotion recognition.'' in
  \emph{19th Annual Conference of the International Speech Communication
  Association, INTERSPEECH}, 2018, pp. 3087--3091.

\bibitem{meng2019frame}
D.~Meng, X.~Peng, K.~Wang, and Y.~Qiao, ``Frame attention networks for facial
  expression recognition in videos,'' in \emph{2019 IEEE International
  Conference on Image Processing (ICIP)}.\hskip 1em plus 0.5em minus
  0.4em\relax IEEE, 2019, pp. 3866--3870.

\bibitem{fan2018video-based}
Y.~Fan, J.~C. Lam, and V.~O. Li, ``Video-based emotion recognition using
  deeply-supervised neural networks,'' in \emph{Proceedings of the 20th ACM
  International Conference on Multimodal Interaction}, 2018, pp. 584--588.

\bibitem{55503fc645ce0a409eb301e1}
S.~Poria, E.~Cambria, A.~Hussain, and G.-B. Huang, ``Towards an intelligent
  framework for multimodal affective data analysis,'' \emph{Neural Networks},
  vol.~63, pp. 104--116, 2015.

\bibitem{dobrivsek2013towards}
S.~Dobri{\v{s}}ek, R.~Gaj{\v{s}}ek, F.~Miheli{\v{c}}, N.~Pave{\v{s}}i{\'c}, and
  V.~{\v{S}}truc, ``Towards efficient multi-modal emotion recognition,''
  \emph{International Journal of Advanced Robotic Systems}, pp. 1--10, 2013.

\bibitem{song2004audio}
M.~Song, J.~Bu, C.~Chen, and N.~Li, ``Audio-visual based emotion recognition -
  {A} new approach,'' in \emph{2004 {IEEE} Computer Society Conference on
  Computer Vision and Pattern Recognition {(CVPR} 2004), with CD-ROM, 27 June -
  2 July 2004, Washington, DC, {USA}}, 2004, pp. 1020--1025.

\bibitem{zeng2006training}
Z.~Zeng, Y.~Hu, M.~Liu, Y.~Fu, and T.~S. Huang, ``Training combination strategy
  of multi-stream fused hidden markov model for audio-visual affect
  recognition,'' in \emph{Proceedings of the 14th ACM international conference
  on Multimedia}, 2006, pp. 65--68.

\bibitem{fan2016video}
Y.~Fan, X.~Lu, D.~Li, and Y.~Liu, ``Video-based emotion recognition using
  cnn-rnn and c3d hybrid networks,'' in \emph{Proceedings of the 18th ACM
  International Conference on Multimodal Interaction}, 2016, pp. 445--450.

\bibitem{li2019bi}
S.~Li, W.~Zheng, Y.~Zong, C.~Lu, C.~Tang, X.~Jiang, J.~Liu, and W.~Xia,
  ``Bi-modality fusion for emotion recognition in the wild,'' in \emph{2019
  International Conference on Multimodal Interaction}, 2019, pp. 589--594.

\bibitem{zhou2019exploring}
H.~Zhou, D.~Meng, Y.~Zhang, X.~Peng, J.~Du, K.~Wang, and Y.~Qiao, ``Exploring
  emotion features and fusion strategies for audio-video emotion recognition,''
  in \emph{2019 International Conference on Multimodal Interaction}, 2019, pp.
  562--566.

\bibitem{zhang2017learning}
S.~Zhang, S.~Zhang, T.~Huang, W.~Gao, and Q.~Tian, ``Learning affective
  features with a hybrid deep model for audio--visual emotion recognition,''
  \emph{IEEE Transactions on Circuits and Systems for Video Technology},
  vol.~28, no.~10, pp. 3030--3043, 2017.

\bibitem{li2019attentive}
J.-L. Li and C.-C. Lee, ``Attentive to individual: A multimodal emotion
  recognition network with personalized attention profile.'' in \emph{20th
  Annual Conference of the International Speech Communication Association,
  INTERSPEECH}, 2019, pp. 211--215.

\bibitem{zhang2019deep}
Y.~Zhang, Z.-R. Wang, and J.~Du, ``Deep fusion: An attention guided factorized
  bilinear pooling for audio-video emotion recognition,'' in \emph{2019
  International Joint Conference on Neural Networks (IJCNN)}.\hskip 1em plus
  0.5em minus 0.4em\relax IEEE, 2019, pp. 1--8.

\bibitem{2018Audiovisual}
E.~Avots, T.~Sapi{\'n}ski, M.~Bachmann, and D.~Kami{\'n}ska, ``Audiovisual
  emotion recognition in wild,'' \emph{Machine Vision and Applications},
  vol.~30, no.~5, pp. 975--985, 2019.

\bibitem{hu2017learning}
P.~Hu, D.~Cai, S.~Wang, A.~Yao, and Y.~Chen, ``Learning supervised scoring
  ensemble for emotion recognition in the wild,'' in \emph{Proceedings of the
  19th ACM international conference on multimodal interaction}, 2017, pp.
  553--560.

\bibitem{liu2018multi}
C.~Liu, T.~Tang, K.~Lv, and M.~Wang, ``Multi-feature based emotion recognition
  for video clips,'' in \emph{Proceedings of the 20th ACM International
  Conference on Multimodal Interaction}, 2018, pp. 630--634.

\bibitem{wang2019multi}
Y.~Wang, J.~Wu, and K.~Hoashi, ``Multi-attention fusion network for video-based
  emotion recognition,'' in \emph{2019 International Conference on Multimodal
  Interaction}, 2019, pp. 595--601.

\bibitem{guidi2015automatic}
A.~Guidi, N.~Vanello, G.~Bertschy, C.~Gentili, L.~Landini, and E.~P. Scilingo,
  ``Automatic analysis of speech f0 contour for the characterization of mood
  changes in bipolar patients,'' \emph{Biomedical Signal Processing and
  Control}, vol.~17, pp. 29--37, 2015.

\bibitem{huang2015extraction}
Y.~Huang, A.~Wu, G.~Zhang, and Y.~Li, ``Extraction of adaptive wavelet packet
  filter-bank-based acoustic feature for speech emotion recognition,''
  \emph{IET Signal Processing}, vol.~9, no.~4, pp. 341--348, 2015.

\bibitem{zhao2019speech}
J.~Zhao, X.~Mao, and L.~Chen, ``Speech emotion recognition using deep 1d \& 2d
  cnn lstm networks,'' \emph{Biomedical Signal Processing and Control},
  vol.~47, pp. 312--323, 2019.

\bibitem{sarma2018emotion}
M.~Sarma, P.~Ghahremani, D.~Povey, N.~K. Goel, K.~K. Sarma, and N.~Dehak,
  ``Emotion identification from raw speech signals using dnns.'' in
  \emph{Interspeech}, 2018, pp. 3097--3101.

\bibitem{huang2018speech}
J.~Huang, Y.~Li, J.~Tao, Z.~Lian \emph{et~al.}, ``Speech emotion recognition
  from variable-length inputs with triplet loss function.'' in \emph{19th
  Annual Conference of the International Speech Communication Association,
  INTERSPEECH}, 2018, pp. 3673--3677.

\bibitem{liang2019cross}
J.~Liang, S.~Chen, J.~Zhao, Q.~Jin, H.~Liu, and L.~Lu, ``Cross-culture
  multimodal emotion recognition with adversarial learning,'' in \emph{ICASSP
  2019-2019 IEEE International Conference on Acoustics, Speech and Signal
  Processing (ICASSP)}.\hskip 1em plus 0.5em minus 0.4em\relax IEEE, 2019, pp.
  4000--4004.

\bibitem{xu2019learning}
H.~Xu, H.~Zhang, K.~Han, Y.~Wang, Y.~Peng, and X.~Li, ``Learning alignment for
  multimodal emotion recognition from speech,'' in \emph{Interspeech}, 2019,
  pp. 3569--3573.

\bibitem{chatziagapi2019data}
A.~Chatziagapi, G.~Paraskevopoulos, D.~Sgouropoulos, G.~Pantazopoulos,
  M.~Nikandrou, T.~Giannakopoulos, A.~Katsamanis, A.~Potamianos, and
  S.~Narayanan, ``Data augmentation using gans for speech emotion
  recognition.'' in \emph{20th Annual Conference of the International Speech
  Communication Association, INTERSPEECH}, 2019, pp. 171--175.

\bibitem{zhang2019attention}
Z.~Zhang, B.~Wu, and B.~Schuller, ``Attention-augmented end-to-end multi-task
  learning for emotion prediction from speech,'' in \emph{ICASSP 2019-2019 IEEE
  International Conference on Acoustics, Speech and Signal Processing
  (ICASSP)}.\hskip 1em plus 0.5em minus 0.4em\relax IEEE, 2019, pp. 6705--6709.

\bibitem{latif2019direct}
S.~Latif, R.~Rana, S.~Khalifa, R.~Jurdak, and J.~Epps, ``Direct modelling of
  speech emotion from raw speech,'' in \emph{Interspeech}, 2019, pp.
  3920--3924.

\bibitem{sarma2019improving}
M.~Sarma, P.~Ghahremani, D.~Povey, N.~K. Goel, K.~K. Sarma, and N.~Dehak,
  ``Improving emotion identification using phone posteriors in raw speech
  waveform based dnn,'' in \emph{20th Annual Conference of the International
  Speech Communication Association, INTERSPEECH}, 2019, pp. 3925--3929.

\bibitem{articlefacs}
P.~Ekman and W.~V. Friesen, \emph{Facial action coding systems}.\hskip 1em plus
  0.5em minus 0.4em\relax Consulting Psychologists Press, 1978.

\bibitem{knyazev2018leveraging}
B.~Knyazev, R.~Shvetsov, N.~Efremova, and A.~Kuharenko, ``Leveraging large face
  recognition data for emotion classification,'' in \emph{13th IEEE
  International Conference on Automatic Face \& Gesture Recognition (FG
  2018)}.\hskip 1em plus 0.5em minus 0.4em\relax IEEE, 2018, pp. 692--696.

\bibitem{tran2015learning}
D.~Tran, L.~Bourdev, R.~Fergus, L.~Torresani, and M.~Paluri, ``Learning
  spatiotemporal features with 3d convolutional networks,'' in
  \emph{Proceedings of the IEEE international conference on computer vision},
  2015, pp. 4489--4497.

\bibitem{yang2018geometry}
P.~Yang, H.~Yang, Y.~Wei, and X.~Tang, ``Geometry-based facial expression
  recognition via large deformation diffeomorphic metric curve mapping,'' in
  \emph{2018 25th IEEE International Conference on Image Processing
  (ICIP)}.\hskip 1em plus 0.5em minus 0.4em\relax IEEE, 2018, pp. 1937--1941.

\bibitem{choi2018recognizing}
D.~Y. Choi, D.~H. Kim, and B.~C. Song, ``Recognizing fine facial
  micro-expressions using two-dimensional landmark feature,'' in \emph{2018
  25th IEEE International Conference on Image Processing (ICIP)}.\hskip 1em
  plus 0.5em minus 0.4em\relax IEEE, 2018, pp. 1962--1966.

\bibitem{song2019facial}
B.~C. Song, M.~K. Lee, and D.~Y. Choi, ``Facial expression recognition via
  relation-based conditional generative adversarial network,'' in \emph{2019
  International Conference on Multimodal Interaction}, 2019, pp. 35--39.

\bibitem{bai2019disentangled}
M.~Bai, W.~Xie, and L.~Shen, ``Disentangled feature based adversarial learning
  for facial expression recognition,'' in \emph{2019 IEEE International
  Conference on Image Processing (ICIP)}.\hskip 1em plus 0.5em minus
  0.4em\relax IEEE, 2019, pp. 31--35.

\bibitem{wu2019continuous}
S.~Wu, Z.~Du, W.~Li, D.~Huang, and Y.~Wang, ``Continuous emotion recognition in
  videos by fusing facial expression, head pose and eye gaze,'' in \emph{2019
  International Conference on Multimodal Interaction}, 2019, pp. 40--48.

\bibitem{mansoorizadeh2010multimodal}
M.~Mansoorizadeh and N.~M. Charkari, ``Multimodal information fusion
  application to human emotion recognition from face and speech,''
  \emph{Multimedia Tools and Applications}, vol.~49, no.~2, pp. 277--297, 2010.

\bibitem{wang2012kernel}
Y.~Wang, L.~Guan, and A.~N. Venetsanopoulos, ``Kernel cross-modal factor
  analysis for information fusion with application to bimodal emotion
  recognition,'' \emph{IEEE Transactions on Multimedia}, vol.~14, no.~3, pp.
  597--607, 2012.

\bibitem{zeng2008audio}
Z.~Zeng, J.~Tu, B.~M. Pianfetti, and T.~S. Huang, ``Audio--visual affective
  expression recognition through multistream fused hmm,'' \emph{IEEE
  Transactions on multimedia}, vol.~10, no.~4, pp. 570--577, 2008.

\bibitem{zhao2009lipreading}
G.~Zhao, M.~Barnard, and M.~Pietikainen, ``Lipreading with local spatiotemporal
  descriptors,'' \emph{IEEE Transactions on Multimedia}, vol.~11, no.~7, pp.
  1254--1265, 2009.

\bibitem{2006The}
O.~Martin, I.~Kotsia, B.~Macq, and I.~Pitas, ``The enterface'05 audio-visual
  emotion database,'' in \emph{22nd International Conference on Data
  Engineering Workshops (ICDEW'06)}.\hskip 1em plus 0.5em minus 0.4em\relax
  IEEE, 2006, pp. 8--8.

\bibitem{busso2008iemocap}
C.~Busso, M.~Bulut, C.-C. Lee, A.~Kazemzadeh, E.~Mower, S.~Kim, J.~N. Chang,
  S.~Lee, and S.~S. Narayanan, ``Iemocap: Interactive emotional dyadic motion
  capture database,'' \emph{Language resources and evaluation}, vol.~42, no.~4,
  pp. 335--359, 2008.

\bibitem{dhall2012collecting}
A.~Dhall, R.~Goecke, S.~Lucey, and T.~Gedeon, ``Collecting large, richly
  annotated facial-expression databases from movies,'' \emph{IEEE Annals of the
  History of Computing}, vol.~19, no.~03, pp. 34--41, 2012.

\bibitem{krizhevsky2012imagenet}
A.~Krizhevsky, I.~Sutskever, and G.~E. Hinton, ``Imagenet classification with
  deep convolutional neural networks,'' \emph{Advances in neural information
  processing systems}, vol.~25, pp. 1097--1105, 2012.

\bibitem{2019multimodalbst}
S.~S.~S. Mahesh G.~Huddar and V.~S. Rajpurohit, ``Multimodal emotion
  recognition using facial expressions, body gestures, speech, and text
  modalities,'' \emph{International Journal of Engineering and Advanced
  Technology}, vol.~8, 2019.

\bibitem{mirsamadi2017automatic}
S.~Mirsamadi, E.~Barsoum, and C.~Zhang, ``Automatic speech emotion recognition
  using recurrent neural networks with local attention,'' in \emph{2017 IEEE
  International Conference on Acoustics, Speech and Signal Processing
  (ICASSP)}.\hskip 1em plus 0.5em minus 0.4em\relax IEEE, 2017, pp. 2227--2231.

\bibitem{aldeneh2017using}
Z.~Aldeneh and E.~M. Provost, ``Using regional saliency for speech emotion
  recognition,'' in \emph{2017 IEEE international conference on acoustics,
  speech and signal processing (ICASSP)}.\hskip 1em plus 0.5em minus
  0.4em\relax IEEE, 2017, pp. 2741--2745.

\bibitem{glorot2011deep}
X.~Glorot, A.~Bordes, and Y.~Bengio, ``Deep sparse rectifier neural networks,''
  in \emph{Proceedings of the fourteenth international conference on artificial
  intelligence and statistics}.\hskip 1em plus 0.5em minus 0.4em\relax JMLR
  Workshop and Conference Proceedings, 2011, pp. 315--323.

\bibitem{hinton2015distilling}
G.~Hinton, O.~Vinyals, and J.~Dean, ``Distilling the knowledge in a neural
  network,'' in \emph{NIPS Deep Learning and Representation Learning Workshop},
  2015.

\bibitem{king2009dlib}
D.~E. King, ``Dlib-ml: A machine learning toolkit,'' \emph{The Journal of
  Machine Learning Research}, vol.~10, pp. 1755--1758, 2009.

\bibitem{carrier2013fer}
P.-L. Carrier, A.~Courville, I.~J. Goodfellow, M.~Mirza, and Y.~Bengio,
  ``Fer-2013 face database,'' \emph{Universit de Montral}, 2013.

\bibitem{2015Bilinear}
T.-Y. Lin, A.~RoyChowdhury, and S.~Maji, ``Bilinear cnn models for fine-grained
  visual recognition,'' in \emph{Proceedings of the IEEE international
  conference on computer vision}, 2015, pp. 1449--1457.

\bibitem{MCBGaoBZD16}
Y.~Gao, O.~Beijbom, N.~Zhang, and T.~Darrell, ``Compact bilinear pooling,'' in
  \emph{Proceedings of the IEEE conference on computer vision and pattern
  recognition}, 2016, pp. 317--326.

\bibitem{li2017factorized}
Y.~Li, N.~Wang, J.~Liu, and X.~Hou, ``Factorized bilinear models for image
  recognition,'' in \emph{Proceedings of the IEEE International Conference on
  Computer Vision}, 2017, pp. 2079--2087.

\bibitem{yu2017multi}
Z.~Yu, J.~Yu, J.~Fan, and D.~Tao, ``Multi-modal factorized bilinear pooling
  with co-attention learning for visual question answering,'' in
  \emph{Proceedings of the IEEE international conference on computer vision},
  2017, pp. 1821--1830.

\bibitem{HadamardLBP}
J.~Kim, K.~W. On, W.~Lim, J.~Kim, J.~Ha, and B.~Zhang, ``Hadamard product for
  low-rank bilinear pooling,'' in \emph{ICLR}, 2017.

\bibitem{provost2013identifying}
E.~M. Provost, ``Identifying salient sub-utterance emotion dynamics using
  flexible units and estimates of affective flow,'' in \emph{2013 IEEE
  International Conference on Acoustics, Speech and Signal Processing}.\hskip
  1em plus 0.5em minus 0.4em\relax IEEE, 2013, pp. 3682--3686.

\bibitem{wollmer2013lstm}
M.~W{\"o}llmer, M.~Kaiser, F.~Eyben, B.~Schuller, and G.~Rigoll,
  ``Lstm-modeling of continuous emotions in an audiovisual affect recognition
  framework,'' \emph{Image and Vision Computing}, vol.~31, no.~2, pp. 153--163,
  2013.

\bibitem{ChibaNI20}
T.~N. Yuya~Chiba and A.~Ito, ``Multi-stream attention-based {BLSTM} with
  feature segmentation for speech emotion recognition,'' in \emph{21st Annual
  Conference of the International Speech Communication Association,
  INTERSPEECH}, 2020, pp. 3301--3305.

\bibitem{2019Exploringzhao}
Z.~Zhao, Z.~Bao, Y.~Zhao, Z.~Zhang, N.~Cummins, Z.~Ren, and B.~Schuller,
  ``Exploring deep spectrum representations via attention-based recurrent and
  convolutional neural networks for speech emotion recognition,'' \emph{IEEE
  Access}, vol.~7, pp. 97\,515--97\,525, 2019.

\bibitem{LiuLWGGD20}
J.~Liu, Z.~Liu, L.~Wang, Y.~Gao, L.~Guo, and J.~Dang, ``Temporal attention
  convolutional network for speech emotion recognition with latent
  representation,'' in \emph{21st Annual Conference of the International Speech
  Communication Association, INTERSPEECH}, 2020, pp. 2337--2341.

\bibitem{115546}
D.~S. {Pallet}, W.~M. {Fisher}, and J.~G. {Fiscus}, ``Tools for the analysis of
  benchmark speech recognition tests,'' in \emph{International Conference on
  Acoustics, Speech, and Signal Processing (ICASSP)}, vol.~1, 1990, pp.
  97--100.

\bibitem{maaten2008visualizing}
L.~Van~der Maaten and G.~Hinton, ``Visualizing data using t-sne.''
  \emph{Journal of machine learning research}, vol.~9, no.~11, pp. 2579--2605,
  2008.

\bibitem{3264992Wang}
C.~Lu, W.~Zheng, C.~Li, C.~Tang, S.~Liu, S.~Yan, and Y.~Zong, ``Multiple
  spatio-temporal feature learning for video-based emotion recognition in the
  wild,'' in \emph{Proceedings of the 20th ACM International Conference on
  Multimodal Interaction}, 2018, pp. 646--652.

\bibitem{Zhengem2021}
C.~Zheng, C.~Wang, and N.~Jia, ``Emotion recognition model based on multimodal
  decision fusion,'' in \emph{Journal of Physics: Conference Series}, vol.
  1873, no.~1, 2021, p. 012092.

\end{thebibliography}

\vspace{-0.95cm}

\begin{IEEEbiography}[{\includegraphics[width=1in,height=1.25in,clip,keepaspectratio]{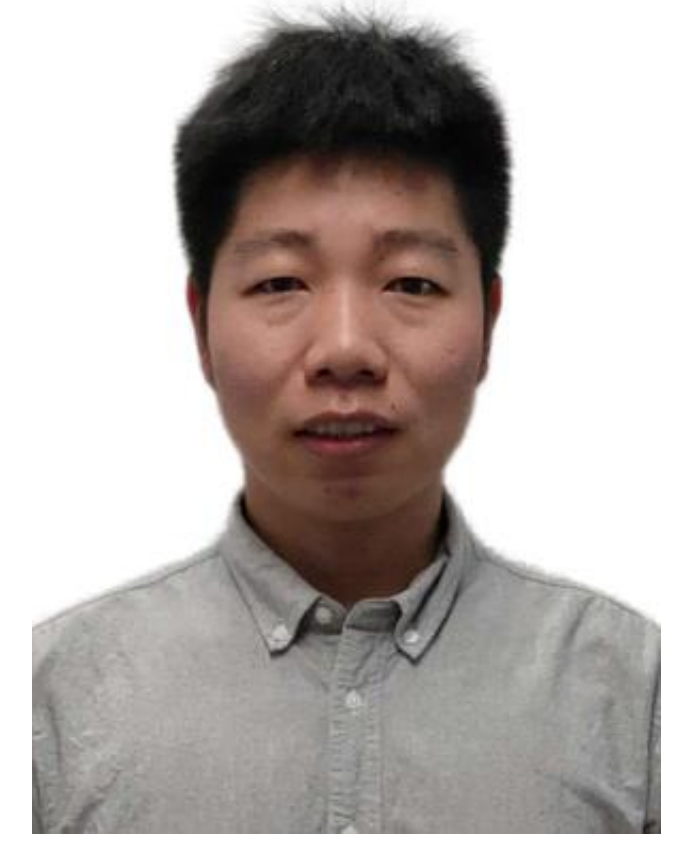}}]{Hengshun Zhou}
received the B.S. degree from the Department of Electronic Information Engineering, Yanshan University, Qinhuangdao, China, in 2016. He is currently working toward the Ph.D. degree with the University of Science and Technology of China (USTC), Hefei, China. His current research mainly includes audio-visual emotion recognition and audio-visual voice activity detection.
\end{IEEEbiography}
\vspace{-0.9cm}
\begin{IEEEbiography}[{\includegraphics[width=1.15in,height=1.25in,clip,keepaspectratio]{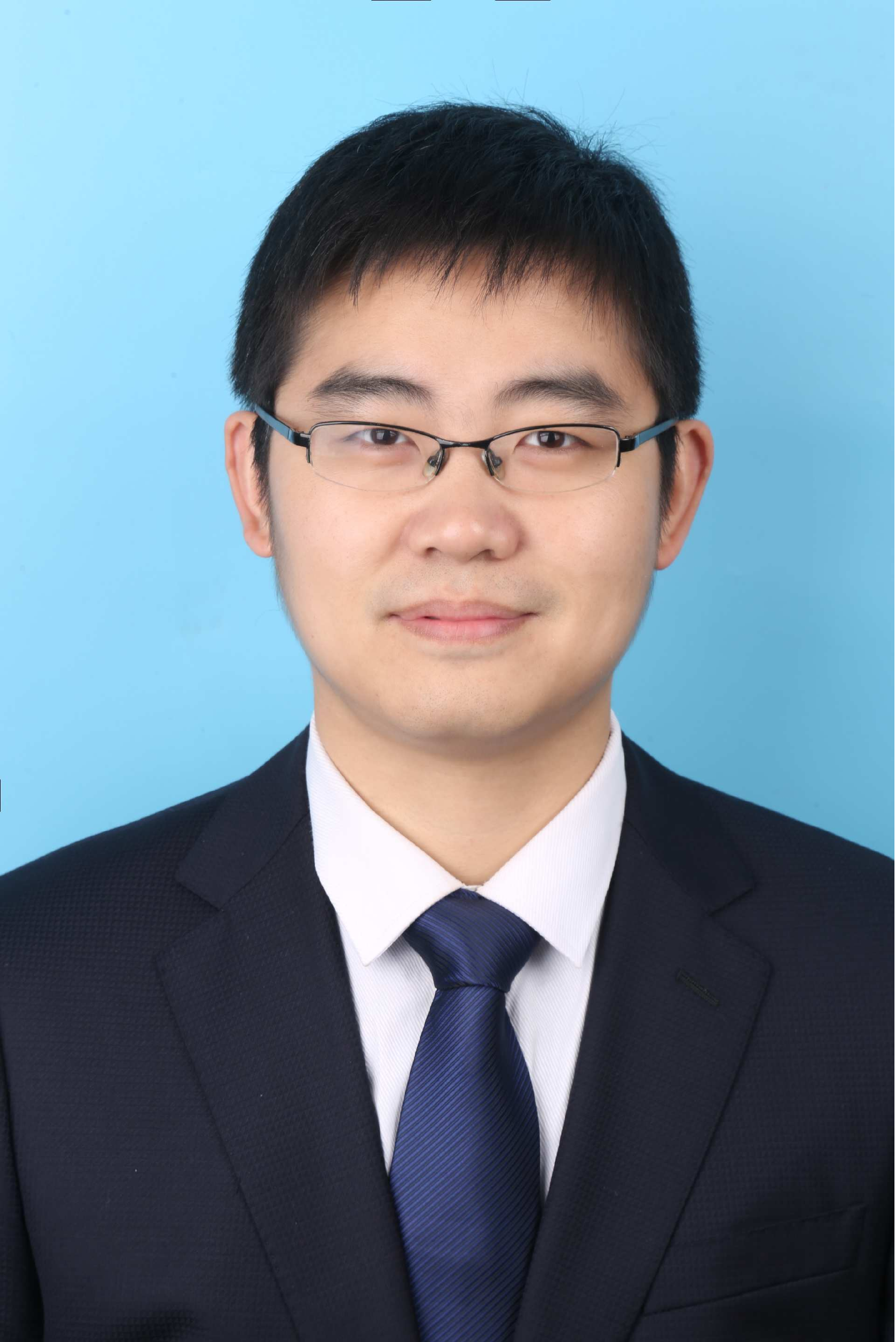}}]{Jun Du}
received the B.Eng. and Ph.D. degrees from the Department of Electronic Engineering and Information Science, University of Science and Technology of China (USTC), in 2004 and 2009, respectively. From July 2009 to June 2010, he worked with iFlytek Research on speech recognition. From July 2010 to January 2013, he worked with Microsoft Research Asia as an Associate Researcher, working on handwriting recognition, OCR, and speech recognition.
Since February 2013, he has been with the National Engineering Laboratory for Speech and Language Information Processing, USTC.
\end{IEEEbiography}
\vspace{-0.95cm}
\begin{IEEEbiography}[{\includegraphics[width=1in,height=1.25in,clip,keepaspectratio]{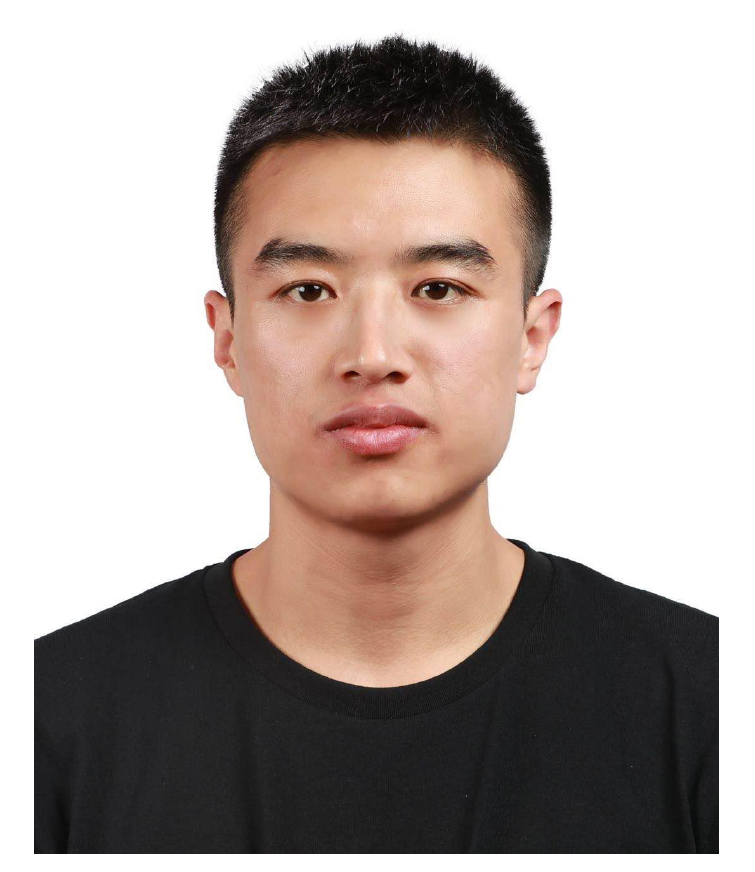}}]{Yuanyuan Zhang}
received the B.S. degree in 2016 and MA.Sc degree in 2019 from the Department of Electronic Engineering and Information Science, University of Science and Technology of China (USTC), Hefei, China. He is currently working in Apple, and his current research mainly involves automatic speech recognition.
\end{IEEEbiography}
\vspace{-0.95cm}
\begin{IEEEbiography}[{\includegraphics[width=1in,height=10.25in,clip,keepaspectratio]{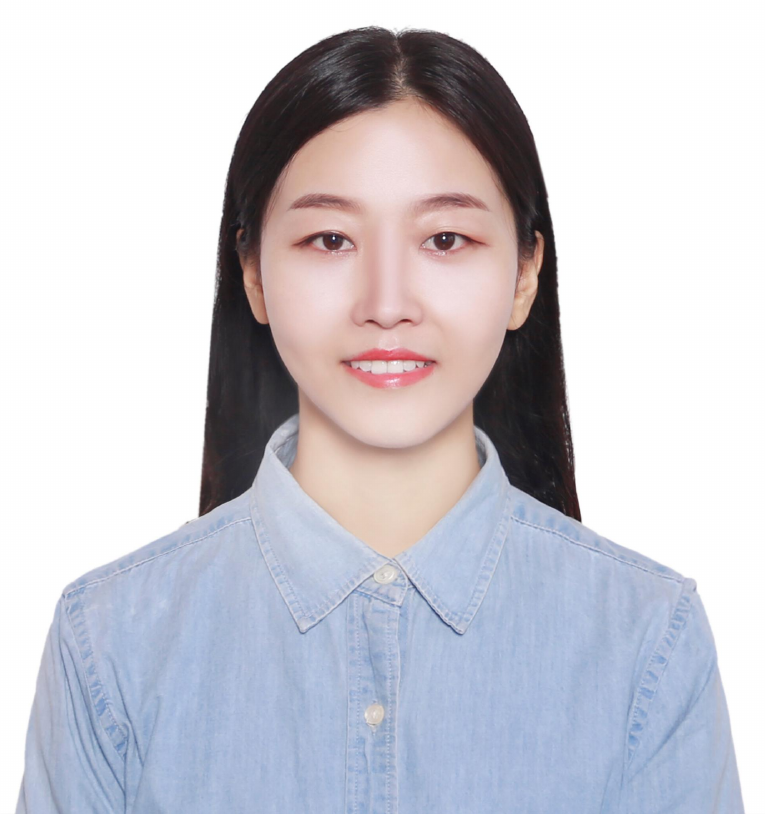}}]{Qing Wang}
received the B.S. degree in 2012 and Ph.D degree in 2018 from the Department of Electronic Engineering and Information Science, University of Science and Technology of China (USTC). She is currently a Post-Doctor with USTC). Her current research interests include speech enhancement, robust speech recognition, and sound event detection.
\end{IEEEbiography}
\vspace{-0.95cm}
\begin{IEEEbiography}[{\includegraphics[width=1in,height=1.25in,clip,keepaspectratio]{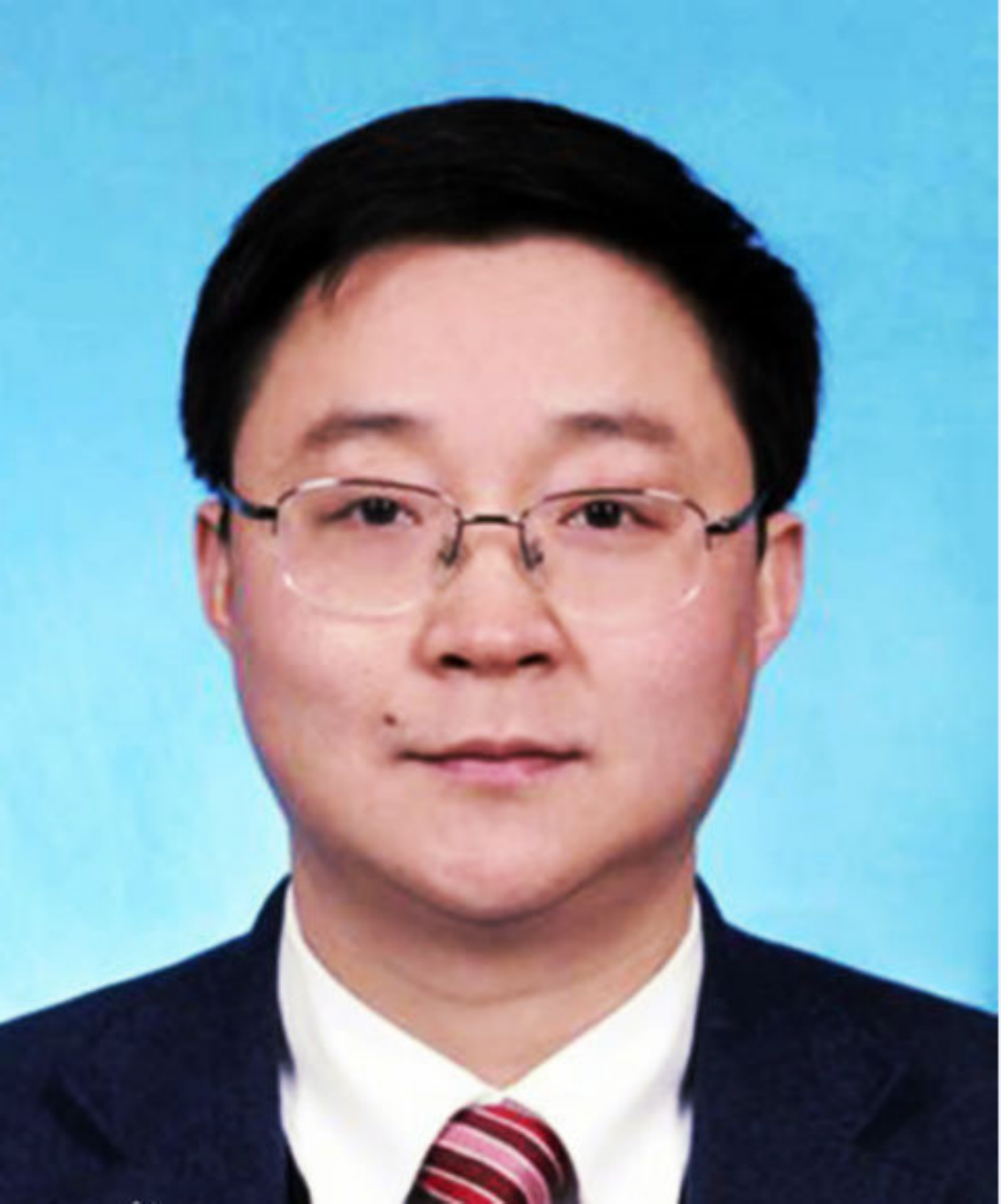}}]{Qing-Feng Liu}
received the B.Eng. and Ph.D. degrees from the Department of Electronic Engineering and Information Science, University of Science and Technology of China (USTC), Hefei, China, in 1998 and 2003, respectively. He is the Founder, CEO, and President of iFLYTEK, the Director of the National Speech and Language Engineering Laboratory of China, a Professor and Doctoral Advisor with USTC, the Director General of the Union of Speech Industry of China, and the Director General of the Union of National University Student Innovation and Entrepreneurship.
\end{IEEEbiography}
\vspace{-0.95cm}
\begin{IEEEbiography}
[{\includegraphics[width=1in,height=1.25in,clip,keepaspectratio]{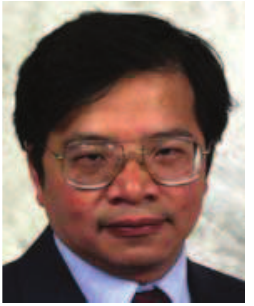}}]{Chin-Hui Lee}
is a Professor with the School of Electrical and Computer Engineering, Georgia Institute of Technology. Before joining academia in 2001, he had 20 years of industrial experience, ending at Bell Laboratories, Murray Hill, NJ, USA, as a Distinguished Member of Technical Staff, and the Director of the Dialogue Systems Research Department. He has authored or coauthored more than 500 papers and 30 patents, and has been cited more than 34 000 times for his original contributions with an h-index of 80 on Google Scholar. He has received numerous awards, including the Bell Labs President¡¯s Gold Award in 1998. He also won SPS¡¯s 2006 Technical Achievement Award for ¡°Exceptional Contributions to the Field of Automatic Speech Recognition.¡± In 2012, he was invited by ICASSP to give a plenary talk on the future of speech recognition. In the same year, he was awarded the ISCA Medal in scientific achievement for pioneering and seminal contributions to the principles and practice of automatic speech and speaker recognition. He is a Fellow of ISCA.
\end{IEEEbiography}


\end{document}